\documentclass[10pt, conference, letterpaper]{IEEEtran}
\IEEEoverridecommandlockouts

\usepackage{cite}
\usepackage{amsmath,amssymb,amsfonts}
\usepackage[T1]{fontenc}
\usepackage{algorithm}
\usepackage{algpseudocode}
\usepackage{amsmath}
\usepackage{amsfonts}
\usepackage{float}

\usepackage{textcomp}
\usepackage{svg}
\usepackage{amsmath}
\usepackage{xcolor}
\usepackage{graphicx}

\usepackage{graphics}
\usepackage{float}
\usepackage{cclicenses}
\usepackage{xspace}
\usepackage{subfigure}
\usepackage{makecell}
\usepackage{multirow} 
\usepackage{wrapfig}
\usepackage{enumitem}
\usepackage{booktabs}
\usepackage{url}
\usepackage{hyperref}
\usepackage{graphicx}      
\usepackage{subfigure}    
\usepackage{makecell,booktabs}

\newcommand{\name}{\texttt{EDT}}
\newcommand{\tpimpr}{$9.5\times$}

\begin{document}
\title{Elastic Data Transfer Optimization with Hybrid Reinforcement Learning}
%
%


\author{\IEEEauthorblockN{Rasman Mubtasim Swargo}
\IEEEauthorblockA{Missouri University of Science and Technology\\
rs75c@mst.edu} 
\and \IEEEauthorblockN{Md Arifuzzaman}
\IEEEauthorblockA{Missouri University of Science and Technology\\ marifuzzaman@mst.edu}
}
\maketitle              

\begin{abstract}
Modern scientific data acquisition generates petabytes of data that must be transferred to geographically distant computing clusters. Conventional tools either rely on preconfigured sessions, which are difficult to tune for users without domain expertise, or they adaptively optimize only concurrency while ignoring other important parameters. We present \name, an adaptive data transfer method that jointly considers multiple parameters. Our solution incorporates heuristic-based parallelism, infinite pipelining, and a deep reinforcement learning based concurrency optimizer. To make agent training practical, we introduce a lightweight network simulator that reduces training time to less than four minutes and provides a $2750\times$ speedup compared to online training. Experimental evaluation shows that \name~consistently outperforms existing methods across diverse datasets, achieving up to \tpimpr~higher throughput compared to state-of-the-art solutions.
\end{abstract}


\maketitle

\section{Introduction}\label{intro}
The modern scientific landscape is experiencing a data deluge, driven by advances in instrumentation, sensing technologies, and computational power that generate datasets at an unprecedented scale. This phenomenon, often referred to as “Big Data,” is commonly characterized by three dimensions: immense volume, high velocity of acquisition, and wide variety of data types. Research infrastructures such as high-energy physics experiments, large-scale genomics initiatives, and global sensor networks now routinely produce petabytes of data. For analysis, this data often must be transferred to geographically distant High-Performance Computing (HPC) clusters~\cite{ligo,ares,combustion,astronomy,des,lsst,atlas,belle2,ameriflux,esgf,esxsnmp}. Moving data at this scale requires extremely high network bandwidth. To address this need, Internet2 has already increased its backbone to 400 Gbps~\cite{internet2400Gbps}, and ongoing efforts aim to extend network capacity to the terabit scale~\cite{esnet1000Gbps}.

To maximize the capabilities of modern transfer infrastructures, it is essential to carefully tune a variety of parameters, such as pipelining (transferring multiple files through a single connection)~\cite{TCP_Pipeline, farkas2002}, parallelism (transferring one file in parts)~\cite{R_Lee01, R_Hacker05, R_Karrer06, R_Lu05} and concurrency (transferring multiple files at a time)~\cite{kosar04, Kosar09, R_Liu10}. These adjustments can be made entirely at the application layer without altering the underlying transfer protocols, and they play a crucial role in enhancing the overall efficiency and throughput of end-to-end data transfers. Most existing approaches to maximizing bandwidth utilization in high-speed networks rely solely on concurrency. This method can be effective for datasets with uniformly sized medium-to-large files. However, it performs poorly when the dataset contains a few very large files or a large number of small files, since it does not take parallelism or pipelining into account.

The key challenge is to design a framework that remains user-friendly, since end users often lack the domain knowledge required to manually tune parameters for optimal performance. Another important consideration is that networks are not static. Bandwidth can fluctuate due to various factors, and a fixed parameter configuration may no longer remain optimal. For this reason, most recent works have modeled the problem as an online optimization task. Hasibul et al.~\cite{drl} addressed this problem using deep reinforcement learning. Although the approach is conceptually sound, it is impractical in practice because it requires day-long online training before the agent can be deployed. 

This paper presents~\textit{\name}, a novel adaptive data transfer method that considers not only concurrency but also pipelining and parallelism. Parallelism is addressed using a heuristic approach based on file size, while pipelining is implemented as an infinite strategy. Concurrency is optimized through a policy-driven deep reinforcement learning (DRL) agent. To eliminate the impracticality of day-long online training~\cite{drl}, we design a lightweight simulator that reduces training time to as little as four minutes.

In summary, the major contributions of this work are:
\begin{itemize}
\item We propose an adaptive multi-parameter framework for maximizing bandwidth utilization during data transfer. Parallelism is determined heuristically, pipelining is implemented as infinite, and concurrency is optimized using a Proximal Policy Optimization (PPO) agent. These three parameters operate transparently, requiring no user expertise. Unlike approaches that rely solely on concurrency, our method generalizes across diverse dataset file-size distributions.

\item We introduce a lightweight network simulator that emulates data transfer dynamics and enables efficient agent training. This simulator achieves a $2750\times$ speedup compared to online training.  

\item We demonstrate through extensive experiments that~\name~outperforms existing methods across all dataset scenarios, achieving up to \tpimpr~higher throughput compared to state-of-the-art solutions.

\end{itemize}

\section{Related Work \& Motivation}\label{motivation}
\textbf{Trade-off Between Performance and Overhead}: Both heuristic and supervised machine learning models estimate parameter values before data transfers begin, using fixed configurations throughout the entire process. This static approach prevents adaptation to changing network conditions during long-running transfers. Additionally, these solutions lack a back-off mechanism, meaning that large parameter values may overwhelm systems in various production environments. For example, if $x$ parallel TCP connections are required to fully utilize the available bandwidth, assigning $x$ connections to every new transfer becomes inefficient when multiple concurrent transfers are initiated. In a scenario with $N$ concurrent transfers, $Nx$ connections would be created, leading to diminished throughput per connection (by a factor of $N$), but significantly increased overhead.
A more efficient alternative would be to allocate fewer connections, say, 10\% of $x$ -- to each new transfer, under the assumption that overlapping transfers or other processes will compete for I/O resources. For instance, Globus Online, a widely adopted data transfer service, allows tuning of three key parameters: pipelining, parallelism, and concurrency. However, in production environments, system administrators set these values heuristically and often very conservatively to minimize I/O and network contention. While this approach helps balance overhead with resource utilization, it frequently leaves networks underutilized. As a result, static configurations present a dilemma: either prioritize utilization at the cost of high overhead, or reduce overhead and risk underutilizing the network.

\textbf{Long Convergence Time of Adaptive Solutions}:  Static solutions become suboptimal when the host or network environments deviate from initial assumption; a common occurrence in long-running transfers due to constantly evolving network dynamics. Moreover, these solutions lack an effective back-off mechanism to manage overhead. Ideally, at the start of a transfer, a system should maximize its bandwidth usage, but as new transfers join, existing transfers should release some connections. Similarly, when a transfer finishes, remaining transfers should immediately capture the freed bandwidth. However, to be useful, these solutions must quickly converge to optimal configurations at the beginning of a transfer or whenever network conditions change, minimizing the search time spent in suboptimal states. The convergence speed is directly related to the search space: with N parameters and M possible ranges per parameter, the search space is $M^N$. This exponential growth in the search space with each additional parameter increases convergence time significantly.

Stochastic approximation algorithms~\cite{rao2016experimental, yun2017data, esma-tcc, liu2018toward} offer promising alternatives to static solutions as they can discover optimal parameter configuration in the runtime. However, Existing  solutions in this domain fall short to offer practical option due to long convergence time and failure to provide fairness and stability guarantees in shared environments. for example, Yun et al. proposed ProbData~\cite{yun2017data} to tune the number of parallel streams and buffer size for memory-to-memory TCP transfers using stochastic approximation. ProbData is able to explore the near-optimal configurations through sample transfers but it takes several hours to converge which makes it impractical to use as majority of the transfers in HPNs last less than few hours~\cite{liu2018cross}. Also, we have observed that background traffic changes drastically over several hours in shared networks~\cite{arslan2018high}, so it may even fail to converge due to large variations in sample transfers. 


\textbf{Simplified Architecture Overwhelm Systems}: The most recent studies also have implemented adaptive approaches using online black-box optimization algorithms such as Bayesian methods, Gradient Descent, and Deep Reinforcement Learning~\cite{arifuzzaman2023falcontpds,marlin,drl,concurrency}. These methods offer high utilization with reduced overhead by periodically collecting socket statistics during transfers. The black-box design allows for instant deployment across various networks, as they don't rely on historical data. These studies finds out that, Concurrency is the most impactful parameter for data transfers. As a result, these solutions narrow the problem to optimizing the number of Concurrency, which can offset suboptimal settings in other parameters. For instance, on a 100 Gbps link, an optimally tuned connection might achieve 10 Gbps, so the solution would require 10 parallel connections. But if TCP buffer size value is suboptimal, throughput per connection could drop to 1 Gbps, prompting the optimizer to increase concurrency to 100. The optimizer can very quickly, often as little in 30 seconds, find this target value due to reduced search space. While these solutions now can easily achieved high network utilization, the increasing concurrency values might significantly degrade the performance of the existing systems and I/O processes.
Although extending these solutions to multi-parameter optimization is fairly straightforward, they often perform much worse because the optimizer spends significant time searching for the optimal configuration due to exponentially increasing search space.

\section{Understanding The Parameters}\label{parameters}
The fundamental challenge in maximizing data transfer throughput over High-Performance Networks (HPNs) is an optimization problem constrained by the Bandwidth-Delay Product (BDP) and the limitations of the TCP protocol. To fully utilize the available bandwidth capacity $Bandwidth_{path}$, the aggregate amount of unacknowledged data in flight must equal the BDP of the network path.
\begin{equation}
    BDP = Bandwidth_{path} \times RTT
\end{equation}
For a single TCP stream, the throughput $T_{single}$ is bounded by the sender's congestion window ($W$) and the Round Trip Time ($RTT$): $T_{single} \le W/RTT$. In modern wide-area networks (WANs), the theoretical BDP often exceeds the operating system's maximum allowed TCP buffer size ($W_{max}$), meaning $W_{max} \ll BDP$. Consequently, a single stream, or even a naive combination of streams, cannot saturate the link, leaving valuable bandwidth wasted.

To achieve $T_{aggregate} \approx Bandwidth_{path}$, the application layer must manipulate three distinct parameters: Pipelining, Concurrency, and Parallelism, to ensure the summation of all active congestion windows approaches the BDP: $\sum W_{i} \approx BDP$.
\begin{figure*}[ht]
    \centering
    \includegraphics[width=0.99\textwidth]{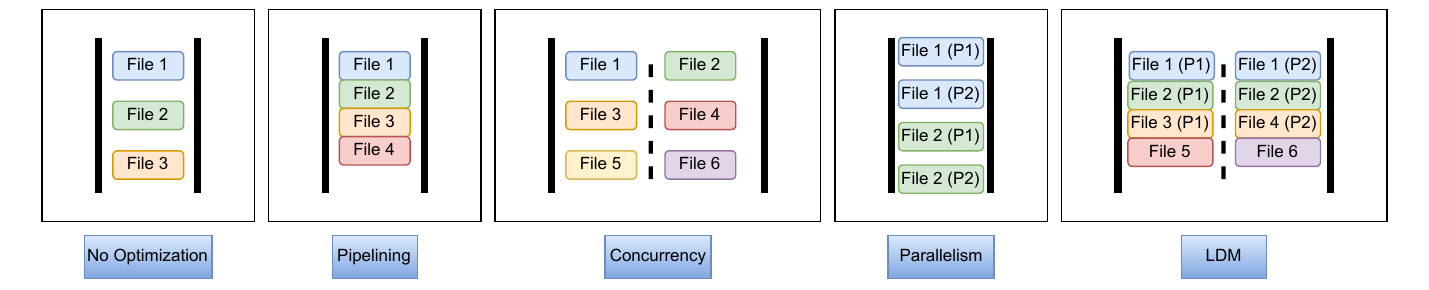}
    \caption{Visual comparison of data transfer behaviors under different optimization strategies. Pipelining eliminates control-channel idle gaps, concurrency transfers multiple files simultaneously, and parallelism splits large files. \name~combines all three to maximize throughput.}
    \label{fig:explanation}
\end{figure*}

\subsection{Pipelining: Necessity for The Small Files}
When transferring datasets containing large numbers of small files, the primary bottleneck shifts from bandwidth to latency. For each file transfer, standard protocols undergo a control channel exchange to open the data channel, followed by the TCP 3-way handshake and the slow-start phase. If the transmission time of the file is shorter than the control overhead (typically $>1$ RTT), the data channel remains idle for significant periods, as shown in the "No Optimization" column of Figure \ref{fig:explanation}.

\begin{figure}[ht]
    \centering
    \includegraphics[width=0.48\textwidth]{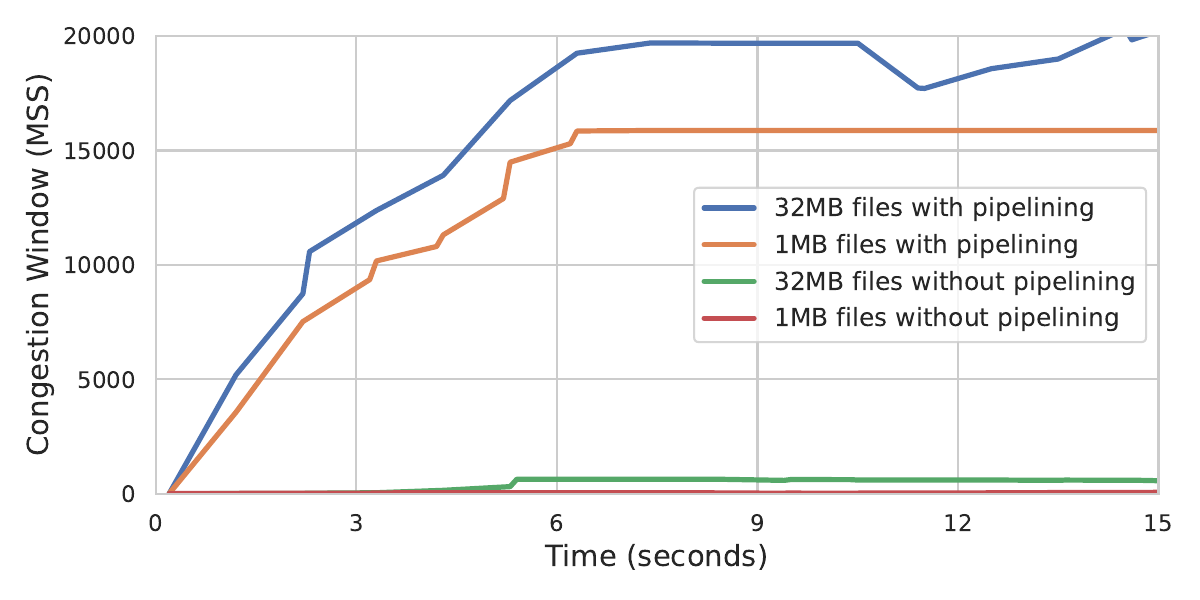}
    \caption{Growth of the TCP congestion window over time for 1 MB and 32 MB files, with and without pipelining in a 35 Gbps - 67 ms RTT testbed.}
    \label{fig:pipelining}
\end{figure}

Pipelining addresses this by treating a sequence of small files as a continuous stream, sending transfer commands back-to-back without waiting for intermediate acknowledgments. The impact of this is quantified in Figure \ref{fig:pipelining}. Without pipelining (Green and Red lines), the TCP connection is torn down and re-established for every file, causing the congestion window to reset repeatedly. Pipelining prevents the TCP window size from shrinking to zero due to idle data channel time. As seen in Figure \ref{fig:pipelining} (Blue and Orange lines), pipelining maintains a persistent data channel, allowing the congestion window to grow continuously until it reaches the network limit, regardless of individual file sizes.

\subsection{Concurrency: Necessary but Insufficient}
Concurrency ($cc$) increases the number of simultaneous file transfers, allowing the application to aggregate the throughput of multiple TCP streams: $T_{agg} = \sum_{i=1}^{cc} T_{i}$. This is the primary mechanism for filling the BDP when $W_{max}$ limits individual streams. 

\begin{figure}[ht]
    \centering
    \includegraphics[width=0.48\textwidth]{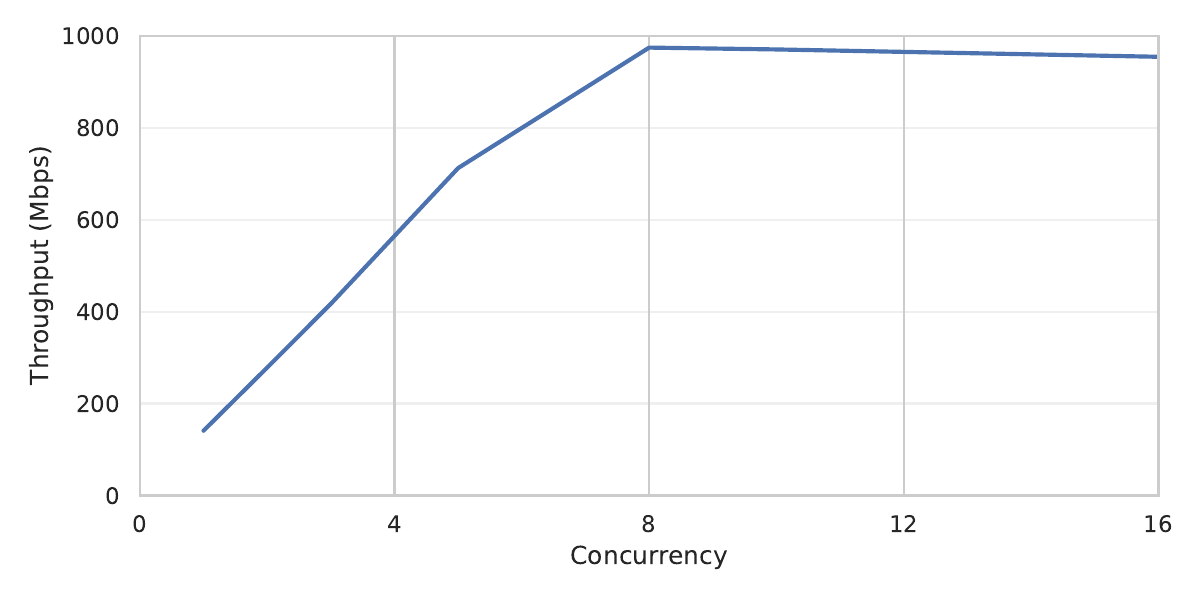}
    \caption{Impact of increasing concurrency on aggregate throughput. Throughput scales with the number of concurrent transfers until system and network bottlenecks begin to limit additional gains.}
    \label{fig:concurrency}
\end{figure}

Figure \ref{fig:concurrency} demonstrates the linear scaling provided by concurrency. By adding more concurrent flows, the total volume of data in flight increases, effectively filling the available bandwidth pipe. But unnecessarily high concurrency can introduce overheads such as network congestion, packet loss, and high I/O contention, which ultimately reduce overall throughput. This is why the throughput begins to decline after the concurrency level exceeds 8.

However, if a dataset contains a wide range of file sizes, concurrency alone cannot fully optimize the transfer. As shown in the "Concurrency" column of Figure \ref{fig:explanation}, multiple files may be transferred at once, but overheads begin to dominate when the files are much smaller than the BDP. 

\subsection{Parallelism: Partitioning Large Files}
While concurrency allows multiple files to share the link, it is strictly bounded by the file count ($N$). If the dataset contains a few very large files ($N < CC_{optimal}$), concurrency alone leaves the network underutilized because there are not enough distinct files to create the necessary number of streams. 
Parallelism improves throughput by splitting large files into multiple chunks and transferring these chunks in parallel. Parallelism is advantageous specifically when the system buffer size is smaller compared to the BDP.

\begin{figure}[ht]
    \centering
    \includegraphics[width=0.48\textwidth]{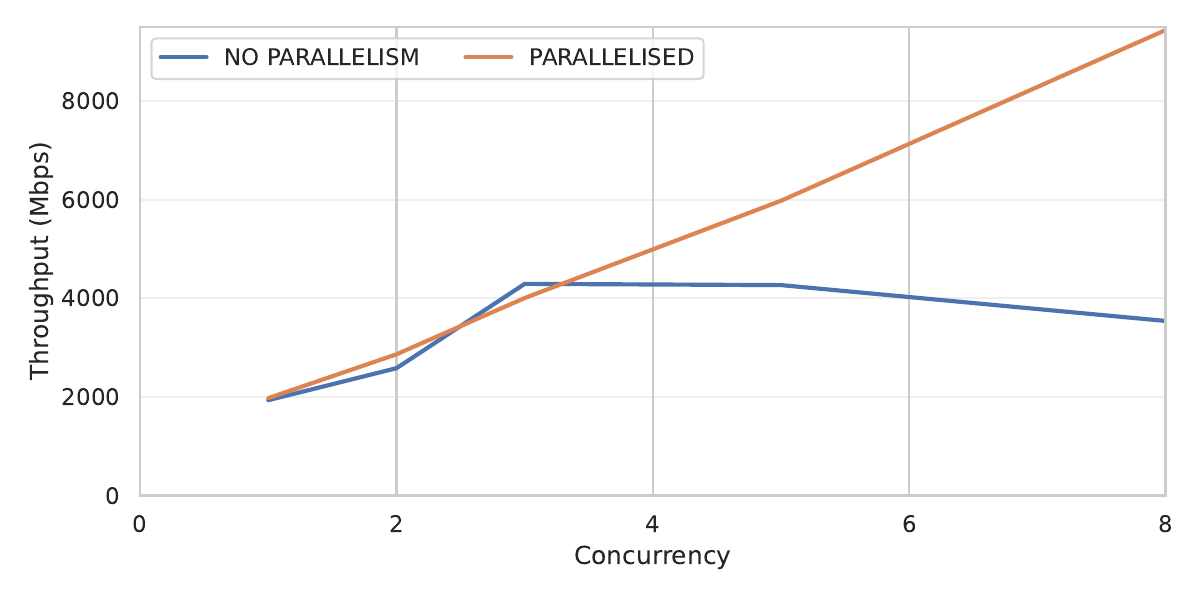}
    \caption{Throughput measured while increasing concurrency for a dataset containing three 2 GB files without parallelism.}
    \label{fig:parallelism}
\end{figure}

Figure \ref{fig:parallelism} demonstrates this effect. When transferring three 2 GB files without parallelism, throughput increases only up to a concurrency level of three. Beyond this point, additional threads have nothing to do, and the throughput plateaus at roughly 4.3 Gbps. By introducing parallelism, each file is divided into multiple chunks (such as P1 and P2 in Figure \ref{fig:explanation}), effectively creating additional "virtual" files. These chunks allow the system to activate more streams, increase the aggregate congestion window, and approach linear throughput scaling. This behavior is visible in the parallelized portion of Figure \ref{fig:parallelism}, where throughput continues to rise as concurrency increases because the dataset now contains many parallel chunks rather than only three files.

\section{\name: Systems Design}\label{ldm}
We introduce \name~to address the throughput maximization problem in high-speed data transfer. It combines both high-level and low-level design strategies. At the high level, we introduce a novel pipeline that determines optimal, stable, and fast-converging transfer parameters. This control strategy is developed and refined using an offline simulator, which allows for efficient agent training. At the low level, we focus on the implementation of this strategy, incorporating robust mechanisms for handling pipelining and parallelism within the production system.

\begin{figure}[ht]
    \centering
    \includegraphics[width=0.47\textwidth]{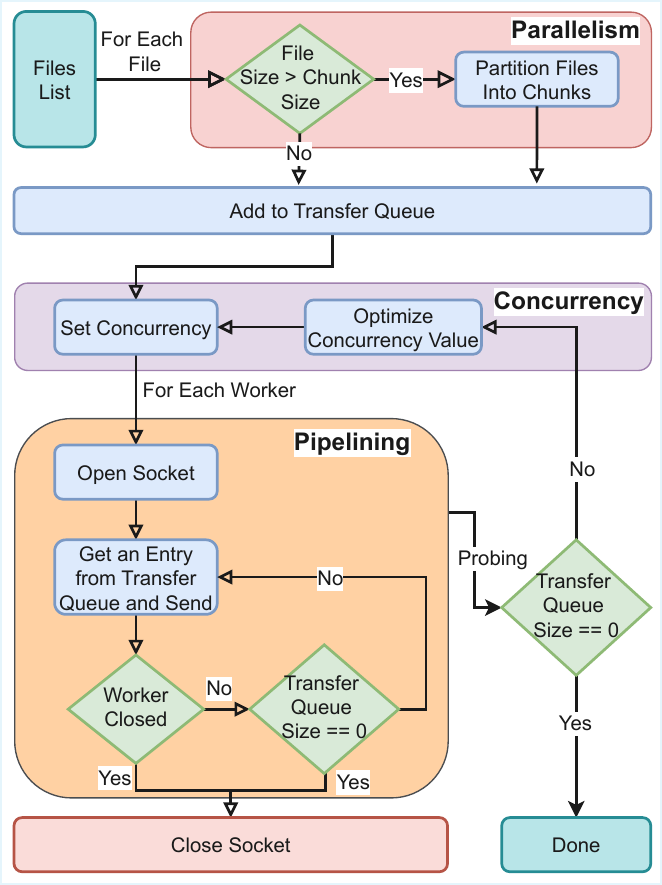}
    \caption{\name~optimizes concurrency, parallelism, and pipelining without being tuned by the end user.}
    \label{fig:diagram}
\end{figure}

\subsection{Handling Transfer Parameters in \name}
In \name, we address the three key transfer parameters—parallelism, pipelining, and concurrency—in unique ways. The overall design is illustrated in the flowchart shown in Figure~\ref{fig:diagram}.

\textbf{Parallelism.} The conventional approach fixes the number of parallel streams per file. However, this method is problematic because it assigns the same number of streams to both a 10 KB file and a 10 GB file. Since parallelism involves splitting and merging files, unnecessary parallelism can create bottlenecks. To overcome this, we implemented a maximum chunk size–based strategy. A maximum chunk size is defined in advance, and each file is transmitted with $\lceil filesize/chunksize \rceil$ parallel streams. This ensures that large files are allocated more streams, while small files typically use only one stream, avoiding unnecessary system overhead.

$chunk size$ is a tunable parameter. If it is set too small relative to the maximum achievable congestion window per stream, each chunk re-enters slow start, creating additional overhead. LDM mitigates this effect through infinite pipelining, which keeps connections warm and reduces the cost of repeated ramp-up, although splitting and merging still introduce their own overhead. Conversely, if the chunk size is too large, streams can experience tail latency because some streams finish early while others remain busy with large residual chunks. In practice, we select a chunk size that balances these effects and prevents pathological fragmentation. In most cases, choosing a chunk size of roughly 10 \% of the network bandwidth (or link capacity) is a safe option because even when tail latency occurs, the remaining chunks complete quickly and do not meaningfully delay the overall transfer.

\textbf{Pipelining.} Pipelining determines how frequently connections are closed and re-established. In \name, we use infinite pipelining: a connection remains open until the worker is stalled by the concurrency controller or the transfer is completed. This approach allows the congestion window to grow to its full potential, avoiding the penalty of restarting from the initial window each time a connection is closed.

\textbf{Concurrency.} Concurrency remains the central and most critical parameter, as both pipelining and parallelism are partially dependent on it. For example, parallelism becomes ineffective when concurrency is fixed at one, and pipelining suffers if concurrency is unstable and frequently turns connections on and off. In \name, concurrency is dynamically determined by a PPO agent, which we describe in detail in the following subsection.

\subsection{Utility Function}
Since we model concurrency as an optimization problem, we require a utility function that captures the essential factors to guide concurrency decisions. For this purpose, we use throughput and packet loss rate as the key parameters. These are collected during the data transfer session and used to assign a score to the current concurrency level.
The utility score increases only when throughput improves meaningfully. Small gains are discouraged, as increasing concurrency for negligible improvement unnecessarily burdens the CPU. For example, suppose concurrency doubles from 8 to 16 but throughput rises only from 20.0 Gbps to 20.5 Gbps. In that case, the utility score penalizes the extra streams since the marginal benefit does not justify the additional resource cost. In essence, the objective is to maximize throughput while minimizing the number of concurrent streams.
The utility function is defined as:

\begin{align*}
U(thrpt, cc) &= \frac{thrpt}{K^{cc}} - B \cdot plr
\end{align*}

Here, $thrpt$ denotes throughput, $cc$ is concurrency, and $plr$ is the packet loss rate, all of which are probed during an ongoing transfer. Throughput contributes positively to the score, while the exponential penalty term $K^{cc}$ discourages excessive concurrency. The constant $K$ is a tunable parameter that balances throughput gains against resource consumption. Similarly, $B$ is a tunable penalty weight that reduces the score in proportion to the packet loss rate, ensuring stability when loss is high.

\subsection{Offline Network Simulator}
To employ a PPO agent as the optimizer, we must first train it. Training such an agent typically requires thousands of episodes, each consisting of multiple steps. This process becomes especially challenging when the reward for an action cannot be obtained immediately. In our case, at least three seconds of transfer data must be observed to compute the parameters of the utility function for a single step. As a result, one episode may take 30 seconds to one minute. Even under optimistic assumptions, no more than 3,000 episodes could be completed in a day, which is insufficient for training a PPO agent effectively. Moreover, in many practical scenarios, intended transfers may last less than one hour, making it impractical to dedicate a full day solely for training.

To address this, we developed an offline simulator capable of reproducing the necessary dynamics in less than four minutes.

\begin{algorithm}
\caption{I/O and Network Dynamics Simulator}
\label{algo:simulator}
\begin{algorithmic}[1]
\State \textbf{Initialization:} Throughputs per thread $TPT$, network bandwidth, initial concurrency, and simulation duration $T_{\text{end}}$.
\medskip
\Function{Task}{$t$, $rem\_bw\_thread$}
    \State $noise \gets random(0,0.8) * MaxFileChunkSize$
    \State $transferSize \gets MaxFileChunkSize - noise$
    \State $throughput\_increase \gets 0$
    \State Task duration, $d_{task} \gets 0$
    \State Compute throughput increase considering overall bandwidth and remaining bandwidth of thread, $rem\_bw\_thread$.
    \State Compute $d_{task}$ according to $TPT$.
    \State Update network throughput and $rem\_bw\_thread$
    \State $t_{\text{next}} \gets t + d_{task} + \epsilon$
    \State \Return $t_{\text{next}}$, $rem\_bw\_thread$
\EndFunction
\medskip
\Function{Get\_Utility}{new\_concurrency}
    \State Reset throughput counter.
    \State Schedule initial tasks and push to transfer queue for each thread in \texttt{new\_threads} with $t = 0$ and $rem\_bw\_thread = TPT$
    \While {the task queue is not empty}
        \State Pop $(t, rem\_bw\_thread)$ from the queue.
        \State $t_{\text{next}}, $rem\_bw\_thread$ \gets$ \Call{Task}{$t, rem\_bw\_thread$}
        \If {$t_{\text{next}} < T_{\text{end}}$}
            \State Add $(t_{\text{next}}, rem\_bw\_thread)$ to the queue.
        \EndIf
    \EndWhile
    \State Normalize throughputs by their finish times.
    \State Compute reward
    \State Update the internal simulator state.
    \State \Return reward and other necessary information.
\EndFunction
\end{algorithmic}
\end{algorithm}

The simulator, described in Algorithm~\ref{algo:simulator}, models all essential aspects of data transfer in a simplified environment. It assumes an infinite pool of files. The files are no larger than the predefined chunk size to ensure parallelism is respected. The simulation is initialized with four parameters: network bandwidth, maximum throughput per thread, initial concurrency, and simulation duration. These values define the environment in which the agent interacts.
Whenever the agent predicts a new action, the simulator triggers a call to the $GET\_UTILITY$ method. Instead of instantiating actual concurrent transfers, the simulator emulates them using queues. Initially, the queue is populated with the current concurrency level. Each element of the queue, representing one concurrency thread, contains the current time and the remaining bandwidth available for that thread during a one-second interval.
Processing proceeds by popping elements from the queue and invoking the \texttt{TASK} method with the time and per-thread bandwidth. This method emulates the time required to send a chunk and returns the next time the thread will begin sending as well as its updated remaining bandwidth. If the next start time exceeds the probing window, the thread is reinserted into the queue. After the queue is fully processed, the throughput is normalized to compute the reward, and the internal state of the simulator is updated accordingly.

\subsection{PPO Agent Architecture}
Policy-driven deep reinforcement learning (DRL) algorithms are well-suited for problems where the environment changes rapidly and a generalizable actor is required. Among these, we adopt Proximal Policy Optimization (PPO)~\cite{schulman2017ppo}, one of the most widely used policy-based reinforcement learning algorithms due to its balance of stability and sample efficiency. Our PPO agent follows the standard PPO architecture, with the key components highlighted in this subsection.

\subsubsection{State Space}
Designing the state space is a crucial step, as it determines how the agent perceives the environment. If irrelevant components are included, the agent may learn misleading correlations with the utility function. At the same time, the state must be expressive enough so that distinct real-world scenarios do not collapse into the same representation. For instance, using only throughput as the state would be problematic, since the same throughput can occur under multiple concurrency settings that require different actions.
Considering these factors, we configure the state with four variables. They are throughput, concurrency, throughput difference from the previous step, and concurrency difference from the previous step. Together, these features define a state representation that can be reliably mapped to an action without confusing the agent.

\subsubsection{Action Space}
In our design, concurrency is selected as the action for the PPO agent. This choice allows the agent to directly map the observed state to the concurrency level, which is the single most important factor in determining the utility value. Other possible design choices do not directly map to the utility; for example, if we define concurrency change as the action, it only modifies concurrency, which in turn affects the utility. Learning this more complex relation requires additional time to converge. 

\subsubsection{Policy Network}
The policy network predicts actions directly from the current state using fully connected layers with residual connections. The input is first embedded into a 256-dimensional space through a linear layer with \texttt{tanh} activation, then passed through three residual blocks. Each block consists of two linear layers with normalization, ReLU, and a skip connection to aid gradient flow and capture complex state representations. The final output is processed by a \texttt{tanh} layer and a linear transformation to compute the mean of the action distribution. A trainable log–standard-deviation parameter, clamped and exponentiated, provides the variance, enabling sampling from a normal distribution to balance deterministic mapping with uncertainty.

\subsubsection{Value Network}
The value network estimates the expected return for a state, essential for PPO advantage calculation. The state is projected into a 256-dimensional feature space using a linear layer with \texttt{tanh} just like the policy network, followed by two residual blocks with Tanh activations and skip connections to preserve gradient flow. The output is then passed through a linear layer to produce a scalar return estimate. This residual-based design improves stability and accuracy in dynamic environments.

\begin{figure}[ht]
    \centering
    \includegraphics[width=0.47\textwidth]{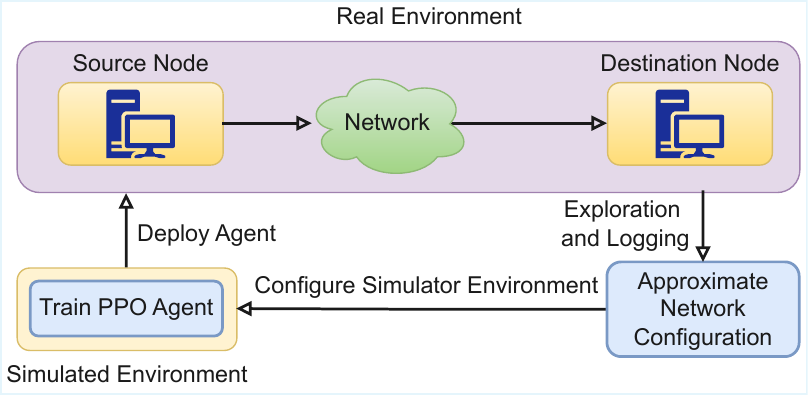}
    \caption{\name~introduces offline training of a deep reinforcement learning agent to quickly learn the behavior of the real environment.}
    \label{fig:workflow}
\end{figure}

\subsection{The \name~Workflow}
After describing the low-level components, we now present the high-level system design of \name, illustrated in Figure~\ref{fig:workflow}. The workflow consists of three phases: the exploration phase, the offline training phase, and the production phase.

\subsubsection{Exploration and Logging}
Before configuring the simulator, we run a short five-minute exploration phase. During this phase, memory-to-memory transfers are executed using random concurrency values. The throughput $T_n$ and concurrency $CC_n$ are logged by probing the network every second, where $n$ denotes the time in seconds.

From these measurements, we estimate the approximate network bandwidth $BW$ and per-concurrency throughput $TPT$ as
\[
BW=\max T_n,\quad
TPT=\max \frac{T_n}{CC_n},\quad
\]
To establish a convergence criterion for the simulator, we assume ideal linear scaling up to the available bandwidth with no packet loss. Under this assumption, the theoretical optimal concurrency $CC^{\star}$ and the maximum achievable reward $R_{\max}$ are defined as
\[
CC^{\star}= \frac{BW}{TPT},\quad
R_{\max}=\frac{BW}{K^{CC^{\star}}}.
\]

\subsubsection{Training in the Simulated Environment}
Once the simulator is configured with the data collected during the exploration phase, we begin training the PPO agent within this simulated environment.

At the start, the policy and value networks are initialized along with an experience buffer. In each episode, the simulator is reset and the agent interacts with the environment for a fixed number of steps. At each step, the policy outputs a Gaussian distribution over concurrency values, from which an action is sampled and executed. The resulting state, reward, and transition are recorded, and episode rewards are accumulated until termination.
After each episode, discounted returns are computed and used to update the networks. The policy is optimized using the clipped surrogate objective to ensure stable learning, while the value network is trained with mean squared error. An entropy regularization term is included to encourage exploration. Both networks are jointly updated using the Adam optimizer, and the old policy is replaced with the updated one.
Training proceeds until convergence is achieved, defined as reaching at least 90\% of the theoretical maximum reward $R_{\max}$. To get the full potential of the agent we keep the training loop running till 1000 episodes pass without any improvement. The best-performing policy and value network are saved as the final model for adaptive thread allocation during transfers.

\subsubsection{Adaptive Concurrency in Production}
In the production phase, we load the best-performing model obtained during training and begin real data transfers. At each step (rather than per episode), the policy network outputs a mean vector $\mu$ and log–standard deviation $\sigma$. An action is sampled from the corresponding diagonal Gaussian distribution and rounded to produce the predicted concurrency $CC_{n+1}$.
\[
CC_{n+1} = \mathrm{round}(\mathcal{N}(\mu, \sigma)).
\]
This value is then clamped to $[1, CC_{\max}]$, where $CC_{\max}$ is a predefined parameter.
Unlike in the training phase, the predicted concurrency is applied directly to stall or reopen workers rather than being passed to the simulator. The network is then probed with the updated concurrency setting to obtain the utility value and the new state. This process repeats continuously until all files in the dataset have been transferred.

\section{Evaluation}\label{eval}
\begin{figure}[t]
  \centering

  \begin{minipage}[b]{0.4\textwidth}
    \centering
    \includegraphics[width=\linewidth]{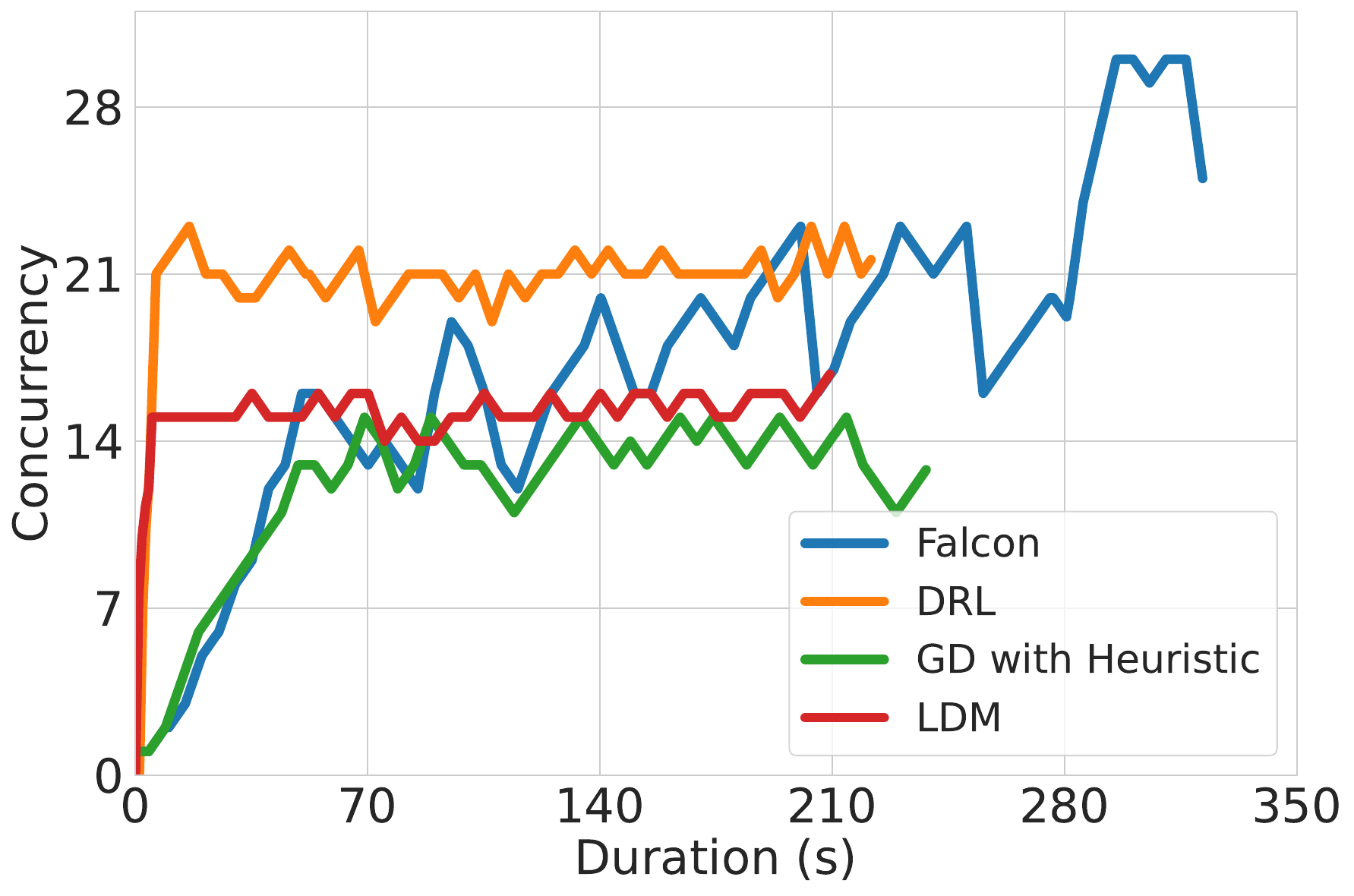}
    \\[-0.3em]
    {\small (a) Concurrency Comparison}
    \vspace{2mm}
  \end{minipage}\hfill
  \begin{minipage}[b]{0.4\textwidth}
    \centering
    \includegraphics[width=\linewidth]{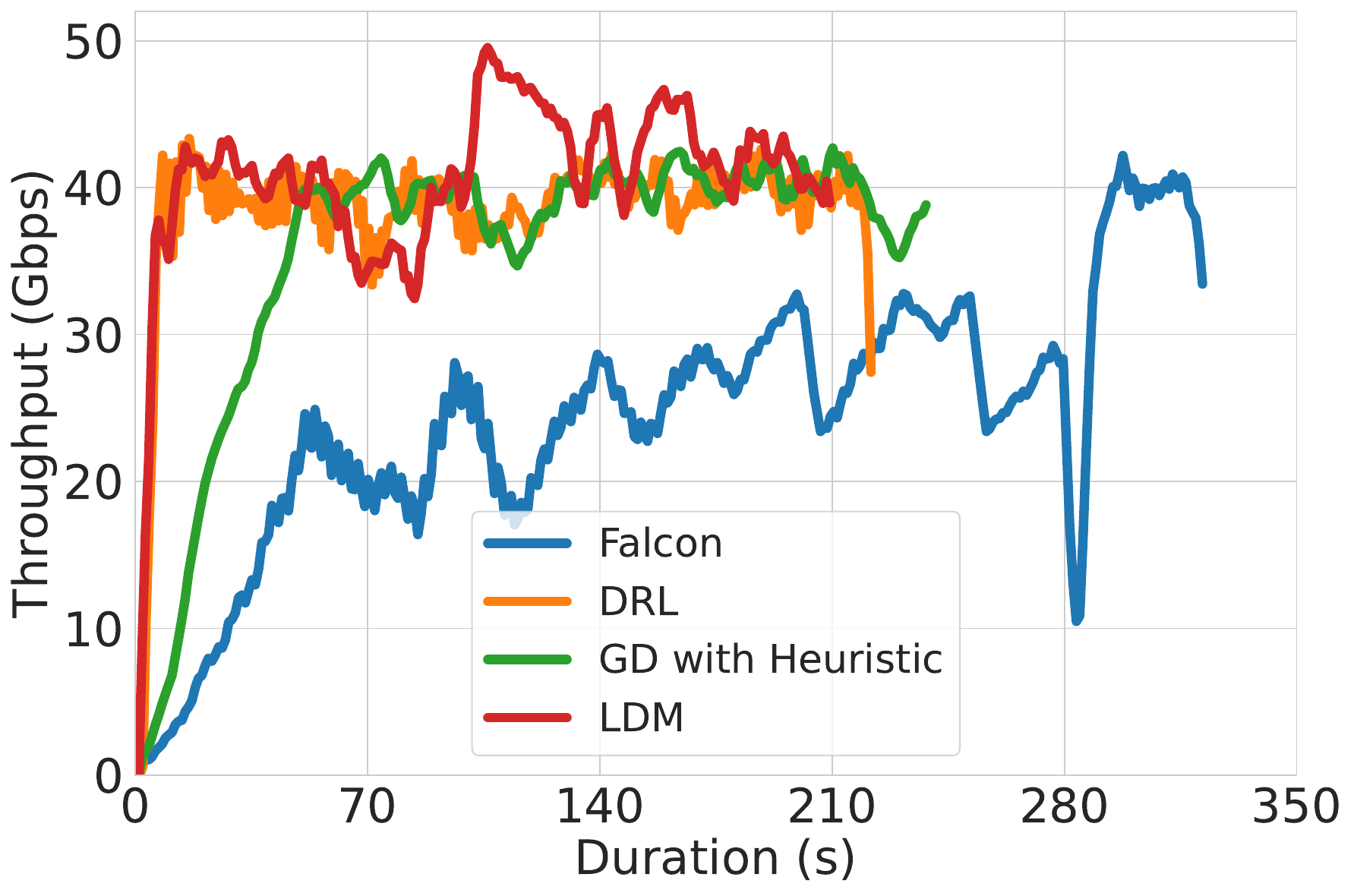}
    \\[-0.3em]
    {\small (b) Throughput Comparison}
  \end{minipage}

  \caption{Performance comparison of \name, Falcon, DRL, and GD with Heuristic with 4096 files sized 256 MB in Fabric-testbed. Only concurrency methods like DRL can reach the top speed, but \name uses 25\% less concurrency.}
  \label{fig:uniform}
\end{figure}

\begin{figure}[t]
  \centering

  \begin{minipage}[b]{0.4\textwidth}
    \centering
    \includegraphics[width=\linewidth]{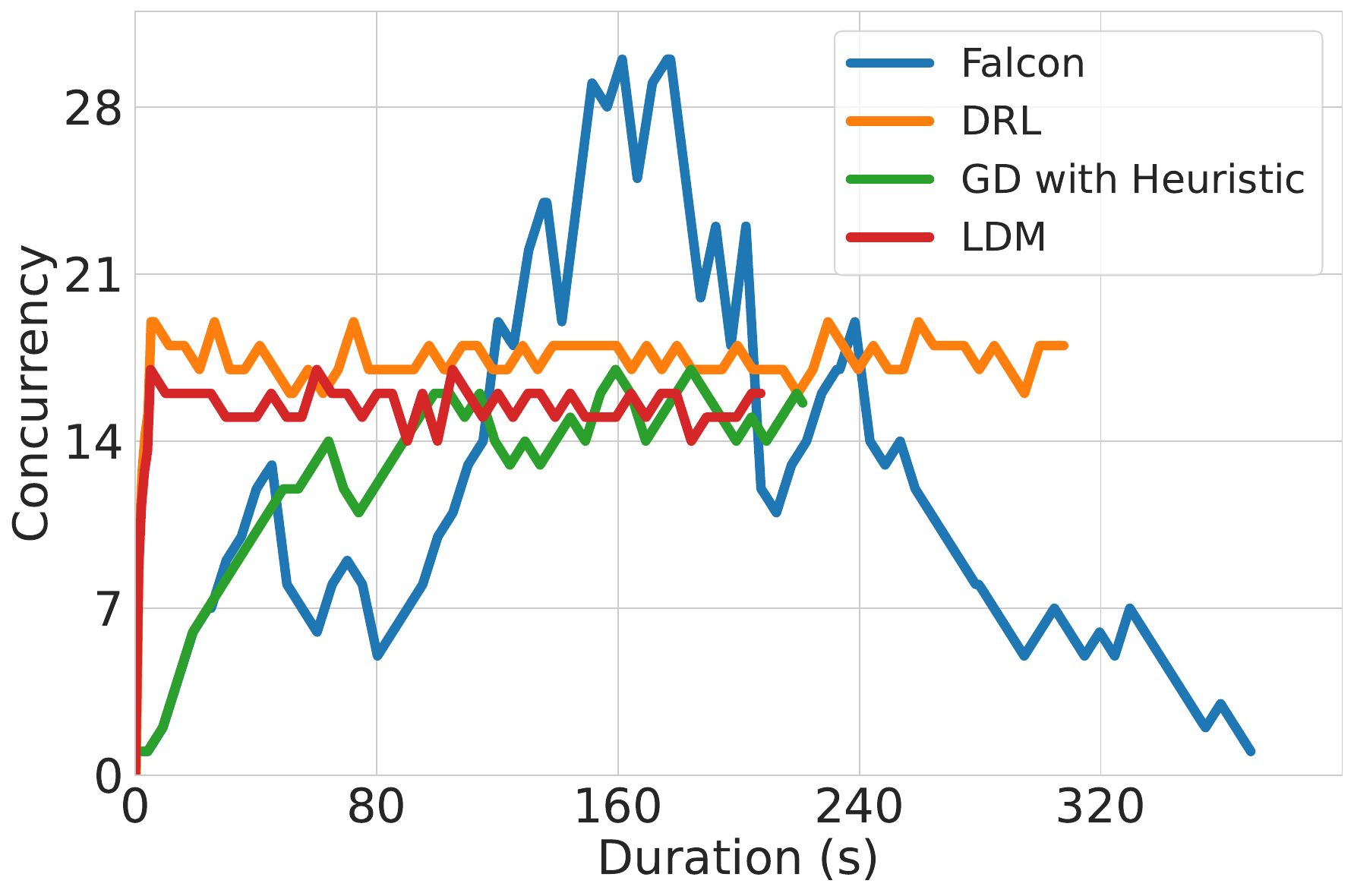}
    \\[-0.3em]
    {\small (a) Concurrency Comparison}
    \vspace{2mm}
  \end{minipage}\hfill
  \begin{minipage}[b]{0.4\textwidth}
    \centering
    \includegraphics[width=\linewidth]{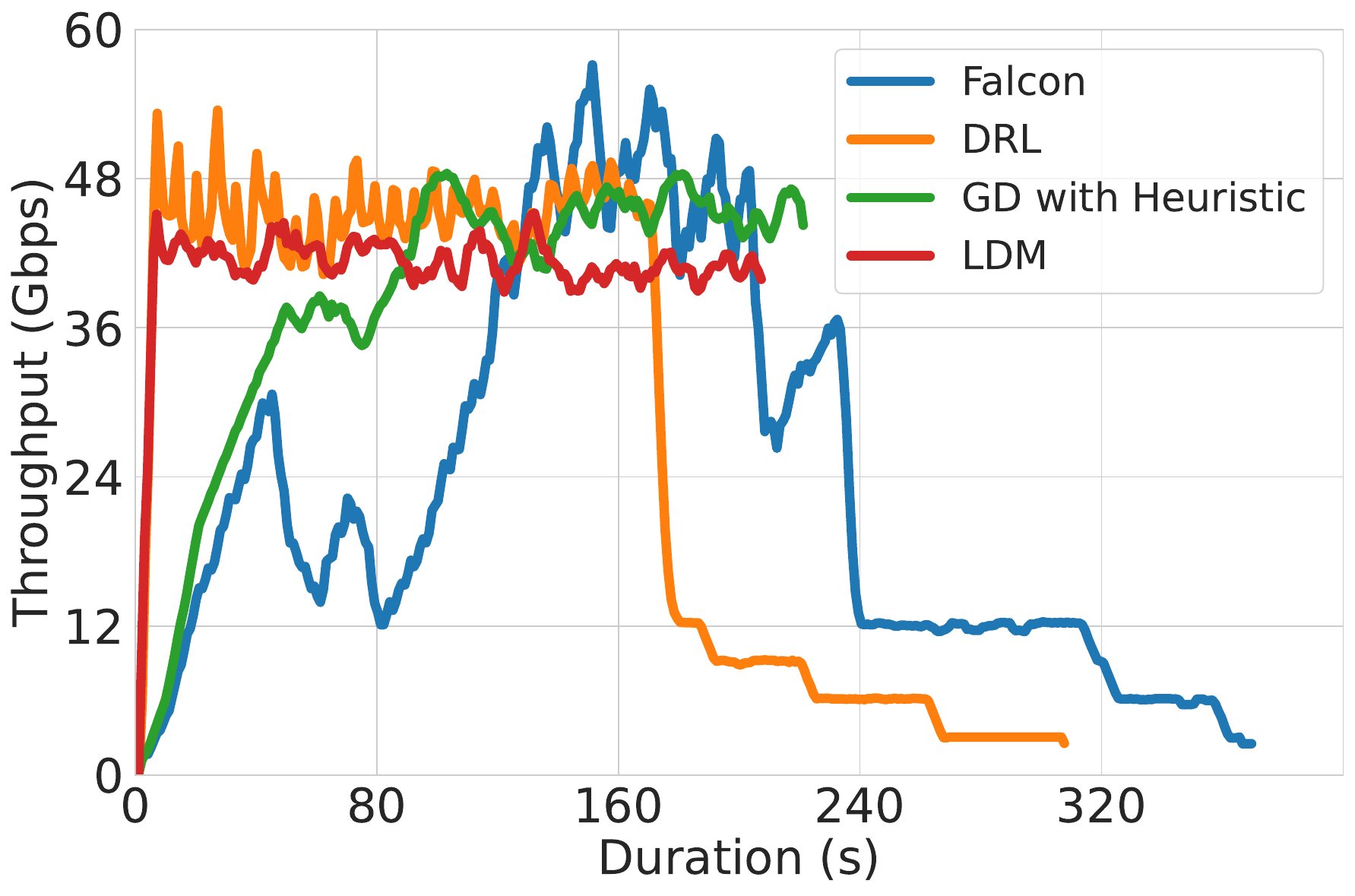}
    \\[-0.3em]
    {\small (b) Throughput Comparison}
  \end{minipage}

  \caption{The tail performance of the methods without parallelism (Falcon and DRL) shows the effect of having 4 large files of 55 GB with 804 files of 1 GB. They cannot utilize the available bandwidth.}
  \label{fig:mid-n-large}
\end{figure}




The goal of this evaluation is to demonstrate the effectiveness, robustness, and practicality of~\name~across a wide range of real-world data transfer scenarios. We first quantify the end-to-end performance gains achieved by~\name~on diverse datasets and network conditions. We then evaluate the generalizability of the system across different TCP congestion control algorithms, followed by an analysis of stability and fairness when multiple transfers share the network. To understand the contribution of each subsystem, we conduct a detailed study of pipelining, parallelism, and concurrency in isolation and in combination. 
Finally, we validate key design choices, including the use of a continuous state space for the reinforcement learning controller. 
Together, these results provide a comprehensive assessment of~\name~and establish its suitability for high-performance data transfer environments. In the experiments described in this section, we used $K = 1.02$, $B = 10$, and $chunksize = 512~MB$.

\begin{table}[ht]
\centering
\caption{Network Testbeds Used in Experiments}
\label{tab:testbeds}
\begin{tabular}{lcccc}
\toprule
\textbf{Testbed} & \textbf{Site 1} & \textbf{Site 2} & \textbf{Observed Bandwidth} & \textbf{RTT} \\
\midrule
Fabric   & NEWY   & CERN   & 45 Gbps & 85 ms \\
Fabric   & MICH   & MASS   & 35 Gbps & 33 ms \\
CloudLab & Clemson & Utah  & 1 Gbps  & 67 ms \\
\bottomrule
\end{tabular}
\end{table}

\subsection{Experimental Setup}
Our evaluation is carried out on two widely used research testbeds, summarized in Table~\ref{tab:testbeds}. The primary experiments are conducted on the Fabric testbed~\cite{fabric-2019} across two geographically distant sites: NEWY (New York, USA) and CERN (Geneva, Switzerland). Although the underlying Fabric links for these sites are provisioned at 100~Gbps and 50~Gbps respectively, Fabric operates as a shared experiment network where Layer~3 routed paths are shared among multiple experiments. As a result, the maximum stable throughputs observed end-to-end are lower than the physical link capacities, and the values reported in Table~\ref{tab:testbeds} reflect these empirically measured stable rates.

Both endpoints on Fabric are provisioned with 8 CPU cores, 64~GB of RAM, a Dell Express Flash P4510 1~TB NVMe SSD, and an NVIDIA Mellanox ConnectX-6 adapter. On the NEWY--CERN path, although the NIC supports a 100~Gbps port rate, the highest stable throughput achievable in practice is approximately 45~Gbps.

To broaden the evaluation across different bandwidth-delay regimes, we also conduct experiments on a second Fabric path between MICH and MASS, which offers a stable throughput of 35~Gbps with a 33~ms RTT. Additionally, to assess~\name~under constrained network conditions, we include experiments on CloudLab between Clemson and Utah, which provides a 1~Gbps link with a 67~ms RTT. Together, these paths span a wide range of bandwidth-delay products, enabling a comprehensive assessment of~\name~across both high-performance and limited-bandwidth environments.
\subsection{Training in the Simulated Environment}
After the logging and exploration phase, we proceed to the offline training of the agent within a simulated environment. We conducted experiments under different scenarios to estimate the time required to satisfy the convergence criteria described in Section~\ref{ldm}. Across 15 experiments, the mean number of episodes to convergence was 8,525 with a standard deviation of 5,065. In terms of wall-clock time, convergence required an average of 156 seconds with a standard deviation of 88 seconds. The maximum observed time was 222 seconds, while the minimum was 33 seconds.

On average, the simulator completed 55 episodes per second. Each episode consists of 10 steps, with each step requiring 5 seconds of probing data~\cite{osti_10324560}. Thus, performing 55 episodes online would take 55 episodes $\times$ 10 steps $\times$ 5 seconds = 2750 seconds. This demonstrates that offline training achieves a speedup of approximately 2,750$\times$ compared to online training.

The simulator preserves the same observable signals as the real network. In practice, it proved highly representative: after deploying the policy and running approximately one hour of online finetuning ($\approx 120$ episodes), the average concurrency changed by only 1 \%. This small adjustment shows that the policy learned in simulation already tracks the real optimal concurrency behavior over time. This negligible improvement makes it impractical for real deployments and is therefore excluded from the final solution.

\begin{table}[ht]
    \centering
    \caption{Dataset Description}
    \label{tab:dataset}
    \renewcommand\arraystretch{1.15}
    \begin{tabular}{@{}lc l@{}}
        \toprule
        \textbf{Dataset} & \textbf{Total Size} & \textbf{File Size Distribution}\\
        \midrule
        A (Uniform) & 1 TB   & \makecell[l]{4096 $\times$ 256 MB} \\
        \midrule
        B (Uniform and Large) & 1 TB 
            & \makecell[l]{804 $\times$ 1 GB \\ 4 $\times$ 55 GB} \\
        \midrule
        C (Small) & 128 GB 
            & \makecell[l]{238{,}770 $\times$ \{100 KB -- 1 MB\}} \\
        \midrule
        D (Mixed) & 256 GB 
            & \makecell[l]{2 $\times$ 12 GB \\ 3872 $\times$ \{10 MB -- 100 MB\} \\ 23{,}624 $\times$ \{100 KB -- 1 MB\}} \\
        \bottomrule
    \end{tabular}
\end{table}

\subsection{End-to-End Performance on Diverse Datasets}
We evaluated \name's performance in terms of throughput maximization, early convergence, and stability across multiple emulated dataset scenarios. For benchmarking, we compared our approach against Falcon~\cite{arifuzzaman2023falcontpds}, a recent state-of-the-art solution; Globus~\cite{globusonline}, the most widely used transfer service; the DRL-based method by Hasibul et al.~\cite{drl}; and a gradient descent optimization strategy with heuristics enabled. In this section, we will refer to Hasibul's method as DRL and the gradient descent optimization strategy with heuristics enabled as GD (Heuristic). Among the evaluated methods, only Globus operates in a static manner, whereas all others are adaptive.
To evaluate the datasets using Globus, we initially attempted benchmarking with Globus Online. However, even after disabling verification, the transfer speed remained very slow. Therefore, for our experiments, we adopted the globus-url-copy utility from the community-maintained open-source Grid Community Toolkit (GCT 6.2).
As discussed in Section~\ref{motivation}, we did not include FDT or MDTM in our comparisons due to the unavailability of their software.

We evaluated \name using the two Fabric testbeds (NEWY–CERN and MICH–MASS) described earlier to capture performance across high-latency and moderate-latency wide-area links. We used BBR as the primary congestion control algorithm because it is specifically designed to operate efficiently in high-bandwidth environments that mirror our target use cases.

To evaluate the generalizability of our approach, we emulated four different datasets, as described in Table~\ref{tab:dataset}. These datasets were specifically designed to expose scenarios where concurrency-based methods are expected to fail due to the absence of pipelining and parallelism.
Each experiment was repeated a minimum of three times to ensure consistency of results.
The experimental results, presented in Tables~\ref{tab:datasets_throughput} and \ref{tab:datasets_throughput_2}, include both mean values (of throughput and concurrency) and their corresponding standard deviations. These results show that~\name~consistently outperforms all baseline methods across diverse dataset configurations.
Across the two Fabric testbeds, \name~achieves speedups of up to 1.02$\times$, 2.83$\times$, 9.58$\times$, and 8.82$\times$ over the state-of-the-art DRL approach.\\

\begin{table*}[t]
\centering
\setlength{\tabcolsep}{4pt} 
\caption{End-to-End Transfer Speed Comparison Across Datasets (Gbps) (Fabric Testbed; NEWY-CERN; BBR congestion Control)}
\label{tab:datasets_throughput}
\begin{tabular}{lcccccccc}
\toprule
\multirow{2}{*}{\textbf{Model}} 
& \multicolumn{2}{c}{\textbf{Dataset A}} 
& \multicolumn{2}{c}{\textbf{Dataset B}} 
& \multicolumn{2}{c}{\textbf{Dataset C}} 
& \multicolumn{2}{c}{\textbf{Dataset D}} \\
\cmidrule(lr){2-3} \cmidrule(lr){4-5} \cmidrule(lr){6-7} \cmidrule(lr){8-9}
 & \textbf{Throughput} & \textbf{Concurrency}
 & \textbf{Throughput} & \textbf{Concurrency}
 & \textbf{Throughput} & \textbf{Concurrency}
 & \textbf{Throughput} & \textbf{Concurrency} \\
\midrule
Globus 
& 4.1 $\pm$ 0.1   & 32.0 $\pm$ 0.0 
& 5.3 $\pm$ 0.2   & 32.0 $\pm$ 0.0 
& 2.5 $\pm$ 0.1   & 64.0 $\pm$ 0.0 
& 5.2 $\pm$ 0.6   & 64.0 $\pm$ 0.0 \\
Falcon 
& 23.7 $\pm$ 1.4 & 16.4 $\pm$ 6.0 
& 22.2 $\pm$ 1.4 & 11.8 $\pm$ 3.5 
& 0.7 $\pm$ 0.1  & 27.6 $\pm$ 13.1 
& 3.4 $\pm$ 0.5  & 14.4 $\pm$ 5.9 \\
DRL 
& 39.5 $\pm$ 0.1 & 20.7 $\pm$ 2.4 
& 24.1 $\pm$ 7.0 & 16.8 $\pm$ 3.3 
& 1.2 $\pm$ 0.5  & 47.5 $\pm$ 2.9 
& 4.0 $\pm$ 0.1  & 16.8 $\pm$ 1.5 \\
\name-GD
& 34.9 $\pm$ 1.7 & 11.7 $\pm$ 3.4 
& 36.4 $\pm$ 3.0 & 12.3 $\pm$ 4.0 
& 11.3 $\pm$ 0.6 & 2.4 $\pm$ 1.1 
& 32.7 $\pm$ 2.5 & 3.8 $\pm$ 2.0 \\
\name~
& \textbf{40.5 $\pm$ 1.2} & 15.1 $\pm$ 1.3 
& \textbf{41.8 $\pm$ 0.2} & 15.5 $\pm$ 1.3 
& \textbf{11.5 $\pm$ 0.3} & 3.8 $\pm$ 0.7 
& \textbf{35.3 $\pm$ 2.7} & 8.6 $\pm$ 0.8 \\
\bottomrule
\end{tabular}
\end{table*}

\begin{table*}[t]
\centering
\setlength{\tabcolsep}{4pt}
\caption{End-to-End Transfer Speed Comparison Across Datasets (Gbps) (Fabric Testbed; MICH-MASS; BBR congestion Control)}
\label{tab:datasets_throughput_2}
\begin{tabular}{lcccccccc}
\toprule
\multirow{2}{*}{\textbf{Model}} 
& \multicolumn{2}{c}{\textbf{Dataset A}} 
& \multicolumn{2}{c}{\textbf{Dataset B}} 
& \multicolumn{2}{c}{\textbf{Dataset C}} 
& \multicolumn{2}{c}{\textbf{Dataset D}} \\
\cmidrule(lr){2-3} \cmidrule(lr){4-5} \cmidrule(lr){6-7} \cmidrule(lr){8-9}
 & \textbf{Throughput} & \textbf{Concurrency}
 & \textbf{Throughput} & \textbf{Concurrency}
 & \textbf{Throughput} & \textbf{Concurrency}
 & \textbf{Throughput} & \textbf{Concurrency} \\
\midrule
Globus 
& 6.2 $\pm$ 0.1   & 32.0 $\pm$ 0.0 
& 3.1 $\pm$ 0.3   & 32.0 $\pm$ 0.0 
& 2.9 $\pm$ 0.2   & 64.0 $\pm$ 0.0 
& 11.3 $\pm$ 0.5   & 64.0 $\pm$ 0.0 \\
Falcon 
& 19.7 $\pm$ 0.9 & 13.6 $\pm$ 5.3
& 9.4 $\pm$ 0.9  & 7.3 $\pm$ 5.2
& 0.96 $\pm$ 0.02  & 16.2 $\pm$ 4.8
& 9.5 $\pm$ 1.0  & 17.8 $\pm$ 5.9 \\

DRL 
& \textbf{38.5 $\pm$ 0.1} & 30.3 $\pm$ 1.2
& 12.4 $\pm$ 0.01 & 30.3 $\pm$ 1.2
& 0.98 $\pm$ 0.01  & 15.3 $\pm$ 0.6
& 10.7 $\pm$ 0.8 & 20.7 $\pm$ 0.9 \\

\name-GD    
& 24.6 $\pm$ 0.2 & 14.9 $\pm$ 5.6
& 24.5 $\pm$ 0.6 & 15.8 $\pm$ 6.2
& 7.7 $\pm$ 0.4  & 2.2 $\pm$ 0.8
& 22.2 $\pm$ 0.2 & 12.2 $\pm$ 3.6 \\

\name~    
& 34.1 $\pm$ 0.3 & 22.9 $\pm$ 0.8
& \textbf{35.1 $\pm$ 0.4} & 27.4 $\pm$ 1.0
& \textbf{7.9 $\pm$ 0.4}  & 2.6 $\pm$ 0.9
& \textbf{26.3 $\pm$ 1.2} & 15.2 $\pm$ 1.8 \\

\bottomrule
\end{tabular}
\end{table*}

\begin{table*}[t]
\centering
\setlength{\tabcolsep}{4pt}
\caption{End-to-End Transfer Speed Comparison Across Datasets (Gbps) (Fabric Testbed; MICH-MASS; CUBIC Congestion Control)}
\label{tab:datasets_throughput_cubic}
\begin{tabular}{lcccccccc}
\toprule
\multirow{2}{*}{\textbf{Model}} 
& \multicolumn{2}{c}{\textbf{Dataset A}} 
& \multicolumn{2}{c}{\textbf{Dataset B}} 
& \multicolumn{2}{c}{\textbf{Dataset C}} 
& \multicolumn{2}{c}{\textbf{Dataset D}} \\
\cmidrule(lr){2-3} \cmidrule(lr){4-5} \cmidrule(lr){6-7} \cmidrule(lr){8-9}
 & \textbf{Throughput} & \textbf{Concurrency}
 & \textbf{Throughput} & \textbf{Concurrency}
 & \textbf{Throughput} & \textbf{Concurrency}
 & \textbf{Throughput} & \textbf{Concurrency} \\
\midrule

Falcon
& 15.6 $\pm$ 0.4 & 12.9 $\pm$ 4.2
& 8.2 $\pm$ 0.2  & 7.3 $\pm$ 5.4
& 1.0 $\pm$ 0.0  & 23.7 $\pm$ 8.1
& 5.2 $\pm$ 1.5  & 11.9 $\pm$ 6.3 \\

DRL
& 21.6 $\pm$ 2.2 & 29.3 $\pm$ 1.4
& 9.2 $\pm$ 0.1  & 13.3 $\pm$ 0.8
& 1.25 $\pm$ 0.0  & 28.7 $\pm$ 1.1
& 7.4 $\pm$ 0.3  & 17.3 $\pm$ 0.7 \\

\name-GD
& 18.3 $\pm$ 0.7 & 17.1 $\pm$ 6.4
& 19.9 $\pm$ 1.3 & 16.43 $\pm$ 6.3
& 7.9 $\pm$ 1.0  & 1.7 $\pm$ 0.8
& 14.9 $\pm$ 0.9 & 7.8 $\pm$ 3.6 \\

\name
& \textbf{23.8 $\pm$ 1.5} & 27.1 $\pm$ 1.7
& \textbf{21.1 $\pm$ 0.7} & 19.5 $\pm$ 1.9
& \textbf{7.4 $\pm$ 0.2}  & 3.2 $\pm$ 0.1
& \textbf{19.7 $\pm$ 0.6} & 10.5 $\pm$ 0.6 \\

\bottomrule
\end{tabular}
\end{table*}


\subsubsection{Effectiveness of Agent-Based Optimization}
Dataset A consists of 4,096 files, each 256 MB in size. This dataset is intentionally designed so that the absence of pipelining and parallelism can be compensated with higher concurrency. This is possible because the number of files is not large enough for pipelining to become a significant bottleneck, and the file sizes are not big enough for parallelism to dominate performance.

To highlight the effectiveness of agent-based methods, we limited the per-concurrency bandwidth to 3 Gbps on the NEWY–CERN path and 1.5 Gbps on the MICH–MASS path. This setup allows us to observe how early convergence and stability influence throughput. 

Globus, by design, is static and operates differently from the other methods considered here. In our approach, parallelism is realized by fixing the chunk size, whereas Globus achieves parallelism by fixing the number of streams per file. In this experiment, Globus was configured with 32 concurrency and 4 parallel streams, resulting in a total of 128 TCP streams. Such a large number of streams, combined with file splitting and merging, can degrade performance. Hence, Globus is expected to underperform in all cases due to its design.

Among adaptive methods, DRL achieves nearly the same throughput as \name, but it does so using $\approx$ 33\% more concurrency in both the testbeds, thereby validating our initial hypothesis.
In the MICH–MASS path, DRL performs better than \name, and this difference can be explained using the bandwidth–delay product (BDP). On the NEWY–CERN path, the maximum achievable congestion window for a single stream is approximately 3000~Mbps (throttled speed) multiplied by 85~ms RTT, which equals 255~Mb, or about 32~MB. On the MICH–MASS path, the corresponding value is 1500~Mbps multiplied by 33~ms RTT, which equals 49.5~Mb, or about 6~MB. In the first case, the file size is roughly eight times larger than the maximum congestion window, while in the second case it is more than forty times larger. As a result, on the MICH–MASS path, the slow-start overhead becomes negligible compared to the additional processing overhead introduced by infinite pipelining and parallelism, making DRL appear faster. However, such homogeneous file distributions are uncommon in real scientific workloads, and the advantage does not generalize beyond this specific scenario.\\
Falcon and GD (Heuristic) perform worse because gradient descent-based online optimization requires additional time to identify the concurrency value that maximizes throughput. As demonstrated in Figure~\ref{fig:uniform}, Falcon and GD (Heuristic) only reached their mean throughput after 46 seconds and 40 seconds, respectively, compared to just 3 seconds and 6 seconds for DRL and \name. This figure illustrates the best runs of each method of the NEWY-CERN path in terms of throughput.

Finally, the standard deviation of concurrency further illustrates the effectiveness of agent-based methods. The standard deviations observed for PPO-based methods (DRL and \name) are less than half of those for their gradient descent counterparts, demonstrating improved stability in optimization.


\subsubsection{Handling Large Files with Parallelism}
We designed Dataset B to include files that would take substantial time to transfer if not split into chunks. The per-concurrency bandwidth was throttled to 3 Gbps on the NEWY–CERN path and 1.5 Gbps on the MICH–MASS path. Specifically, the dataset consists of four 55 GB files along with 804 files of size 1 GB. If the large files were to start simultaneously at the beginning, transferring a single 55 GB file with one thread would take approximately 55 GB * 8 / 3 Gbps $\approx$ 147 seconds for the NEWY-CERN path and 55 GB * 8 / 1.5 Gbps $\approx$ 294 seconds for the MICH-MASS path, even under maximum utilization of that thread. To highlight the necessity of parallelism, we deliberately structured the transfer queue such that the large files were not initiated at the same time. Methods with pipelining can efficiently handle these cases by chunking the large files and distributing them across multiple threads. This phenomenon is illustrated for the NEWY-CERN path in Figure~\ref{fig:mid-n-large}. The figure shows a steep drop in throughput for Falcon and DRL, both of which lack parallelism. Their throughput curves also exhibit a staircase-like pattern at the tail end. In DRL, one large file was completed before the drop, resulting in steps at 9, 6, and 3 Gbps. In Falcon, the steps occurred at 12 and 6 Gbps, indicating that two large files were completed in close succession.
It can be seen that DRL achieves higher throughput than~\name until the medium-sized files finish transferring. This occurs because the processing overhead of~\name, as explained in the previous section, outweighs its benefits when sufficient medium files, which are much larger than the maximum possible congestion window, are still available.

This bottleneck directly impacts overall throughput, as shown in Table~\ref{tab:datasets_throughput} and Table~\ref{tab:datasets_throughput_2}. In this scenario, \name clearly outperforms all baselines, achieving approximately 73\% higher throughput than DRL on the NEWY–CERN path and 273\% higher throughput on the MICH–MASS path. The improvement appears larger on MICH–MASS because the bottleneck files operate at roughly half the per-stream speed (as we bottlenecked) compared to NEWY–CERN, which amplifies the benefits of \name’s optimizations.

\subsubsection{Effect of Pipelining on Small Files}
Dataset C is designed to demonstrate the effect of pipelining. This dataset consists of 238,770 files with sizes ranging from 100 KB to 1 MB. In the earlier cases, the lack of pipelining did not significantly degrade performance because the number of socket closures and reconnections was relatively small. However, with approximately fifty times more files than Dataset A, the overhead of repeatedly closing and reopening sockets becomes the dominant bottleneck.
This effect is evident in the performance comparison shown in Table~\ref{tab:datasets_throughput} and Table~\ref{tab:datasets_throughput_2}. Interestingly, Globus outperforms Falcon and DRL in this scenario. Its fixed concurrency setting provides an advantage, since dynamically adjusting concurrency introduces additional socket operations. In contrast, methods that integrate pipelining are able to overcome this bottleneck and clearly outperform the others.
Nevertheless, the performance of~\name-GD and~\name is somewhat slower than in other datasets, primarily due to the repeated encoding and decoding of file headers during transfers.


\subsubsection{Mixed File Transfers Without Bandwidth Throttling}
The final dataset, Dataset D, was designed to evaluate performance on a mixture of files without throttling per-thread bandwidth. To give methods lacking pipelining a fair opportunity, we reduced the total file count to one tenth of Dataset C. Nevertheless, the results in Table~\ref{tab:datasets_throughput} and Table~\ref{tab:datasets_throughput_2} illustrate that the absence of pipelining still hampers end-to-end throughput.
In this scenario, Falcon and DRL achieved more than a 3$\times$ speedup compared to their performance on Dataset C for both of the testbed paths. However, GD (Heuristic) surpassed DRL by delivering a more than 2$\times$ larger throughput.~\name remained the clear leader, achieving an impressive throughput of 35.3 Gbps (NEWY-CERN) and 26.3 Gbps (MICH-MASS).
\subsection{Generalizability Across TCP Congestion Control Algorithms}
Most recent work on high-performance data transfers adopts BBR because of its model-driven bandwidth probing and resilience to RTT variations. However, CUBIC remains the default congestion control algorithm in Linux systems and is widely deployed on production servers. To ensure that \name’s improvements are not specific to BBR, we repeat all experiments on the MICH–MASS testbed using CUBIC and report the results in Table~\ref{tab:datasets_throughput_cubic}. The trends remain consistent with the BBR experiments in Table~\ref{tab:datasets_throughput_2}.

\name~maintains its relative superiority, outperforming the DRL baseline by approximately 1.1x, 2.3x, 5.9x, and 2.7x across Datasets A, B, C, and D, respectively. For Dataset A, DRL slightly outperformed \name~under BBR because the processing overhead of \name~overshadowed the control-plane bottleneck. Under CUBIC, however, the slower congestion-window growth amplifies control-plane delays, making DRL’s advantage disappear; \name~now leads with 23.76~Gbps versus DRL’s 21.66~Gbps. On Dataset B, where large-file transfers dominate, DRL collapses to 9.16~Gbps due to CUBIC’s conservative window expansion, while \name~reaches 21.09~Gbps through chunk-level parallelism. On Dataset C, pipelining becomes the key factor. Falcon and DRL plateau near 1~Gbps, whereas \name~achieves 7.43~Gbps by continuously issuing file commands and avoiding repeated slow-start resets. For the mixed Dataset D, \name~again delivers the highest throughput at 19.70~Gbps, showing robust adaptation in heterogeneous workloads.

Overall, while CUBIC lowers the achievable throughput ceiling, \name~consistently utilizes a larger fraction of that available bandwidth than all baselines. This confirms that its design generalizes well across both model-based (BBR) and loss-based (CUBIC) congestion control algorithms.

\begin{figure}[t]
  \centering

  \begin{minipage}[b]{0.45\textwidth}
    \centering
    \includegraphics[width=\linewidth]{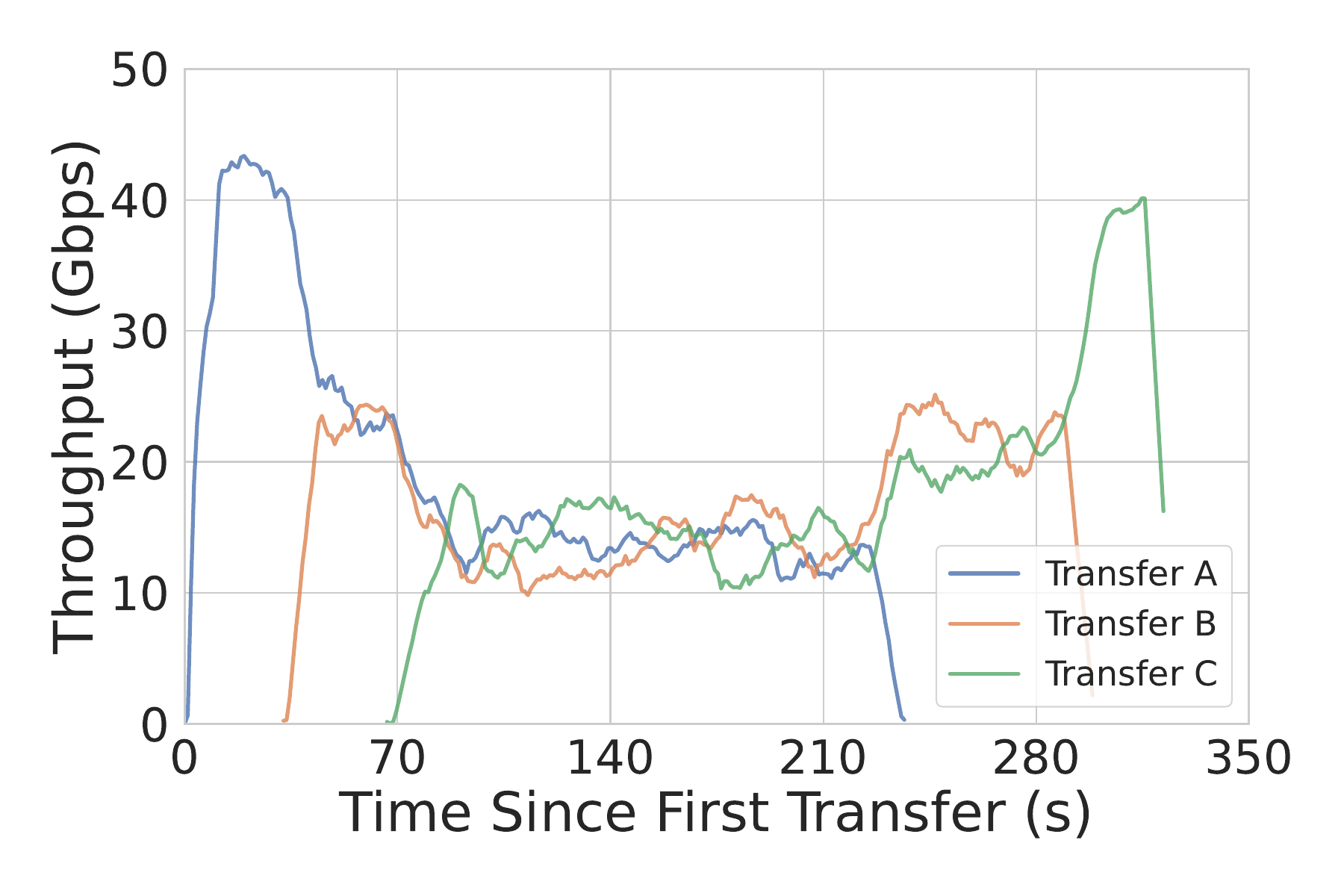}
    \\[-0.3em]
    {\small (a) Instantaneous throughput of competing transfers}
    \vspace{2mm}
  \end{minipage}\hfill
  \begin{minipage}[b]{0.45\textwidth}
    \centering
    \includegraphics[width=\linewidth]{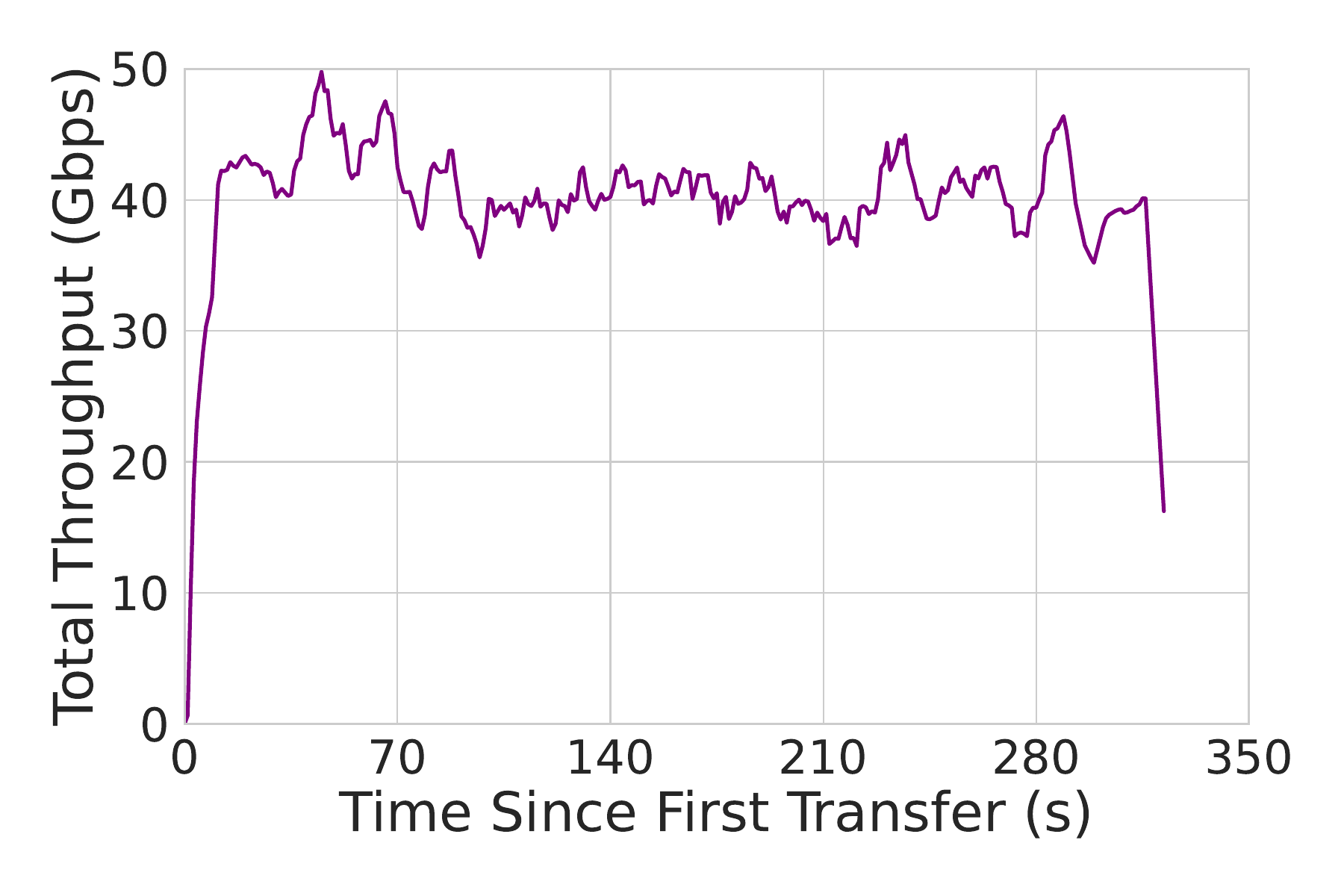}
    \\[-0.3em]
    {\small (b) Aggregate throughput across all transfers}
  \end{minipage}

  \begin{minipage}[b]{0.45\textwidth}
    \centering
    \includegraphics[width=\linewidth]{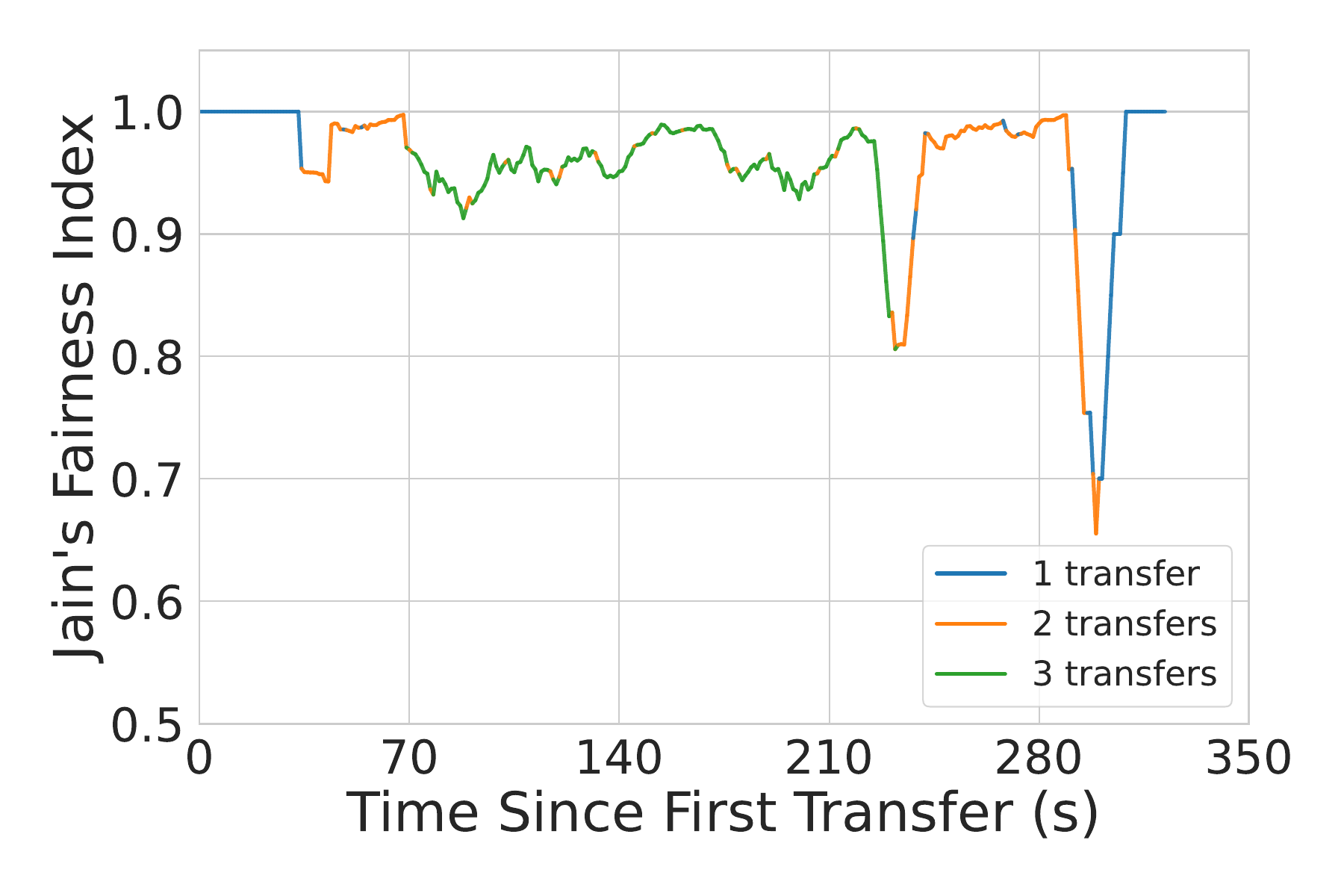}
    \\[-0.3em]
    {\small (c) Jain's Fairness Index}
  \end{minipage}

  \caption{Fairness and link utilization for three competing transfers on the Fabric NEWY-CERN path for Dataset A.}
  \label{fig:competing_transfers}
\end{figure}

\begin{figure}[ht]
    \centering
    \includegraphics[width=0.47\textwidth]{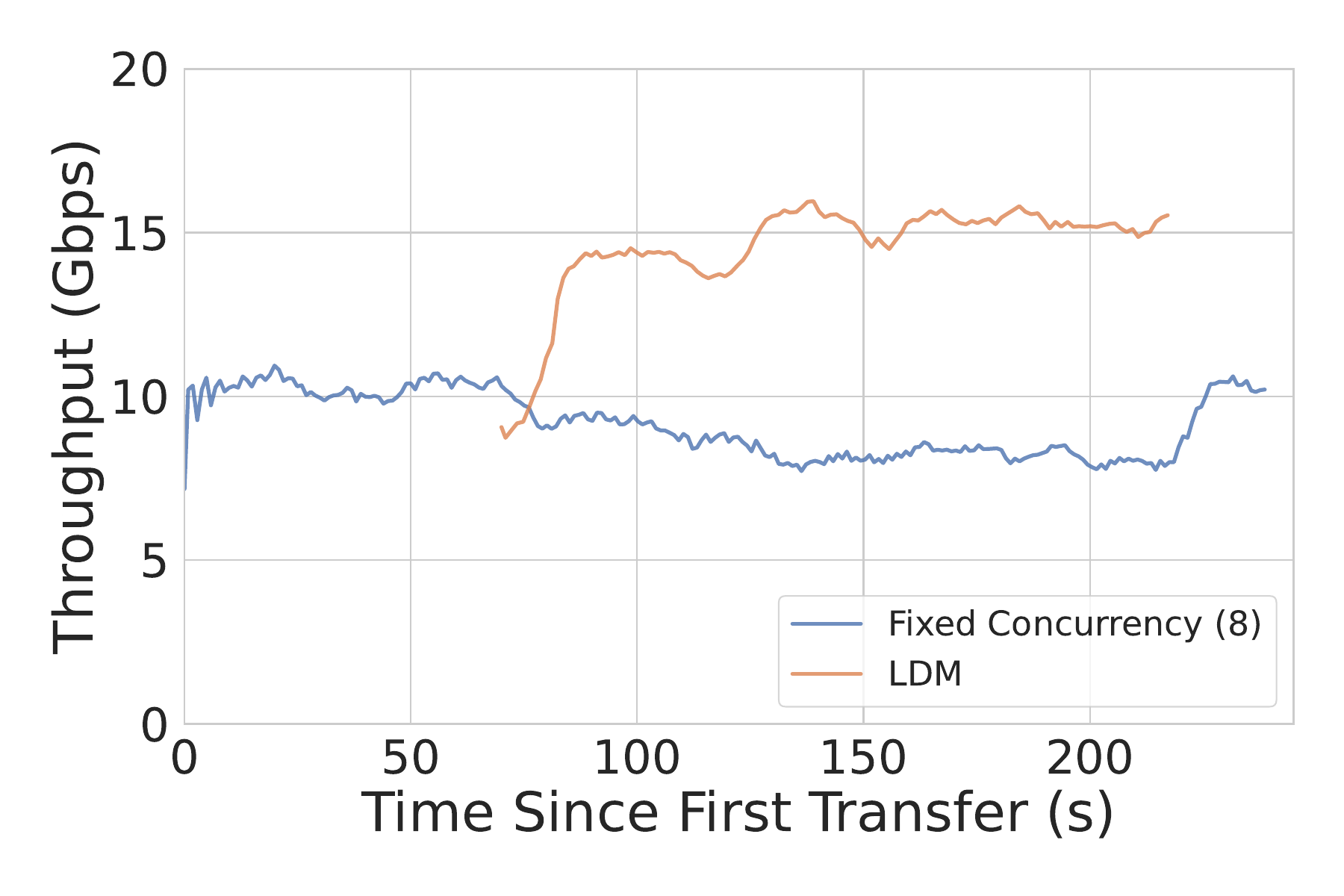}
    \caption{Coexistence of~\name~with a fixed-concurrency cotenant.}
    \label{fig:competing_non_ldm}
\end{figure}

\subsection{Fairness Under Multi-tenant Network Conditions}
In real scientific environments, multiple data transfers often share the same wide-area link, making fairness an essential requirement for practical deployment. To evaluate how~\name~behaves under contention, we run three concurrent transfers on the Fabric NEWY–CERN path and measure their instantaneous and aggregate throughput over time. The results in Figure~\ref{fig:competing_transfers} show that~\name~maintains stable sharing behavior and achieves high overall throughput, unlike methods that rely solely on high concurrency. The aggregated throughput for this experiment using Dataset~A also matches the values reported in Table~\ref{tab:datasets_throughput}, confirming the consistency of our evaluation.
Jain's Fairness Index (JFI) is a standard metric for quantifying how evenly
resources are shared across $n$ flows, defined as:
\[
F = \frac{\left(\sum_{i=1}^{n} x_i\right)^2}{n \sum_{i=1}^{n} x_i^2},
\]
where $x_i$ is the throughput of flow $i$. The index ranges from 0 to 1, with 1 indicating perfectly equal sharing.
Figure~\ref{fig:competing_transfers} (c) shows Jain’s Fairness Index over time for one, two, and three concurrent transfers.
Fairness remains consistently high (typically above 0.95) across all cases, indicating that~\name allocates bandwidth evenly even as new transfers arrive. Small dips occur at the moments when additional transfers join or depart, but the system quickly returns to a stable high-fairness regime.

To evaluate coexistence with heterogeneous cotenants, we additionally run LDM alongside a static transfer configured with fixed concurrency of 8. As shown in Figure~\ref{fig:competing_non_ldm}, the fixed-concurrency flow maintains a stable throughput of roughly 8 Gbps throughout the experiment, while~\name~gradually ramps up and utilizes only the remaining capacity. Even as~\name~converges to higher concurrency values, the non-\name~transfer experiences no noticeable slowdown. This indicates that~\name~plays fair with heterogeneous co-tenants by harvesting spare bandwidth without exhibiting aggressive behavior. Collectively, these results show that LDM achieves high throughput while preserving fairness and coexistence properties in multi-tenant network environments.

\begin{figure}[t]
  \centering

  \begin{minipage}[b]{0.40\textwidth}
    \centering
    \includegraphics[width=\linewidth]{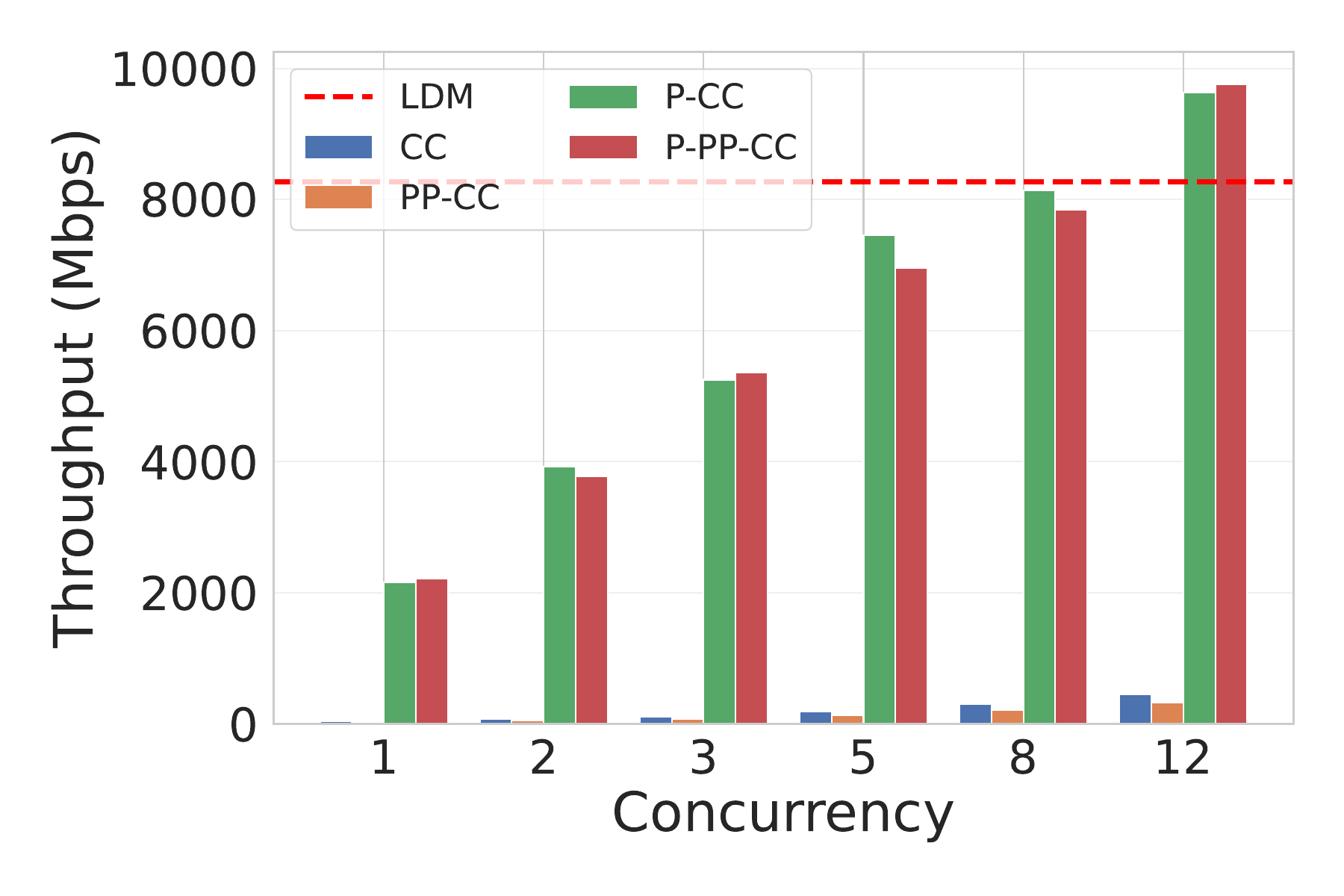}
    \\[-0.3em]
    {\small (a) 5120 files of size 1 MB}
    \vspace{2mm}
  \end{minipage}

  \begin{minipage}[b]{0.40\textwidth}
    \centering
    \includegraphics[width=\linewidth]{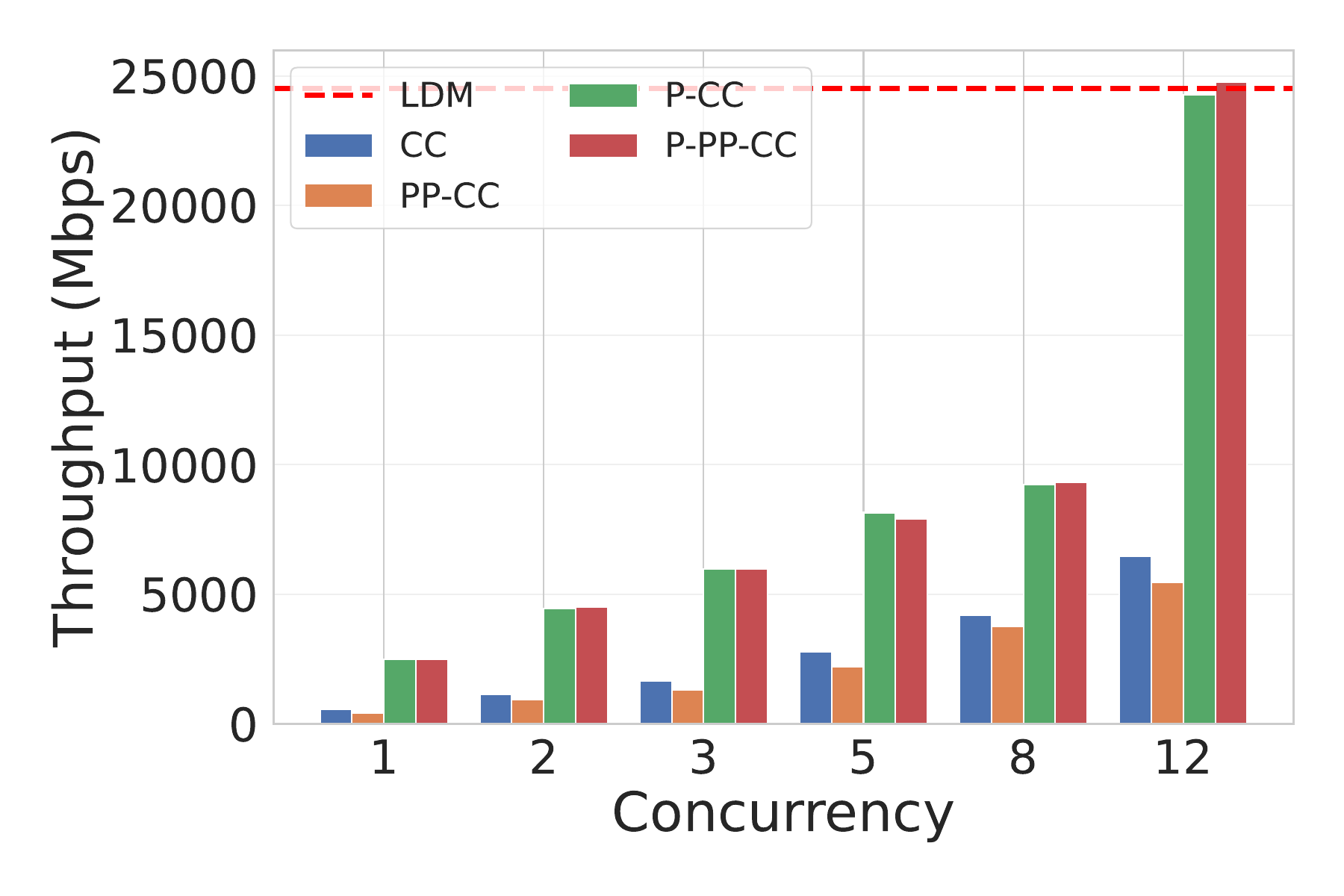}
    \\[-0.3em]
    {\small (b) 160 files of size 32 MB}
    \vspace{2mm}
  \end{minipage}

  \begin{minipage}[b]{0.40\textwidth}
    \centering
    \includegraphics[width=\linewidth]{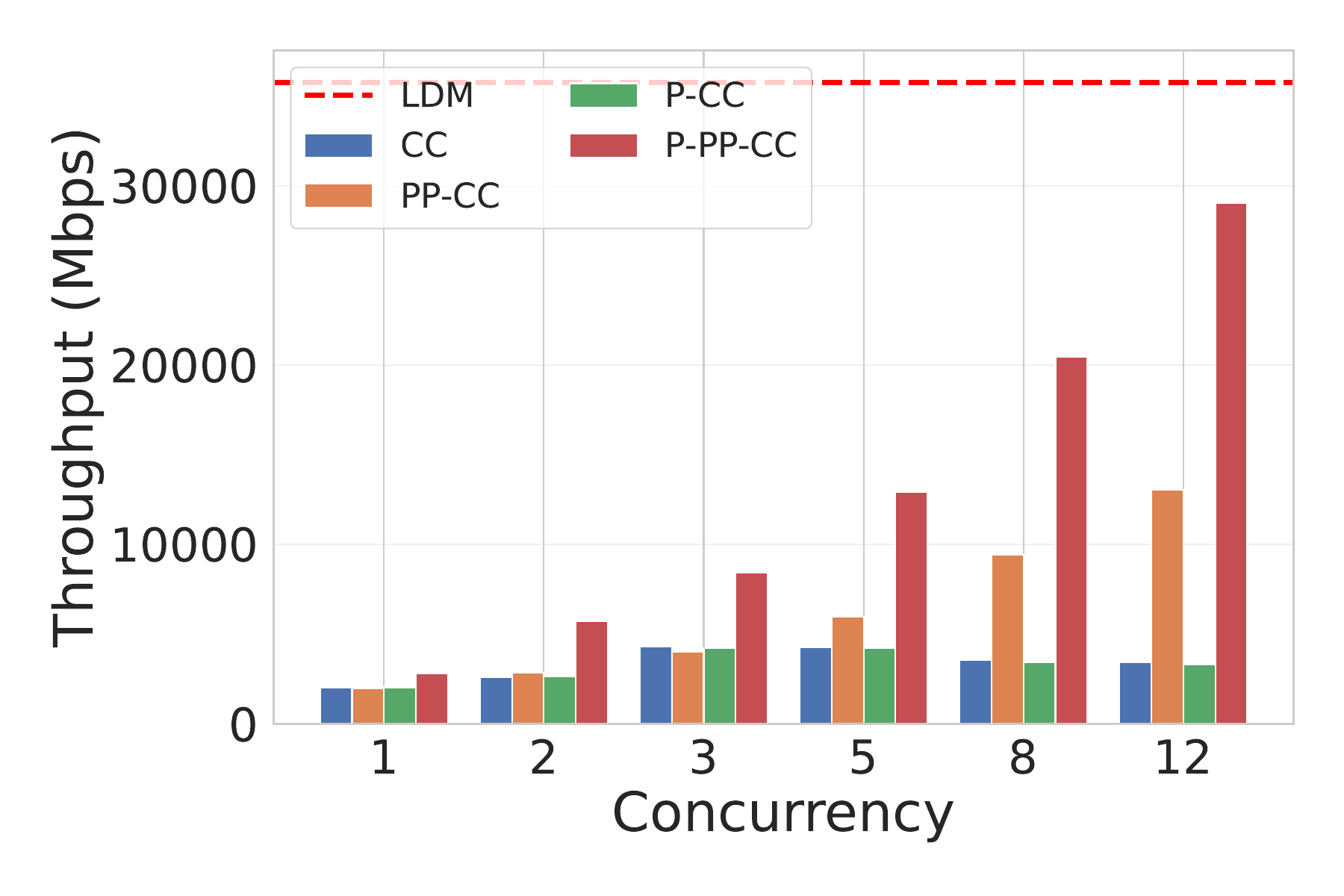}
    \\[-0.3em]
    {\small (c) 3 files of size 2 GB}
  \end{minipage}

  \caption{Ablation study on the Fabric MICH--MASS path for three dataset configurations. Each subplot reports throughput as a function of concurrency for different combinations of pipelining, parallelism, and concurrency.}
  \label{fig:ablation_fabric}
\end{figure}

\begin{figure}[t]
  \centering

  \begin{minipage}[b]{0.40\textwidth}
    \centering
    \includegraphics[width=\linewidth]{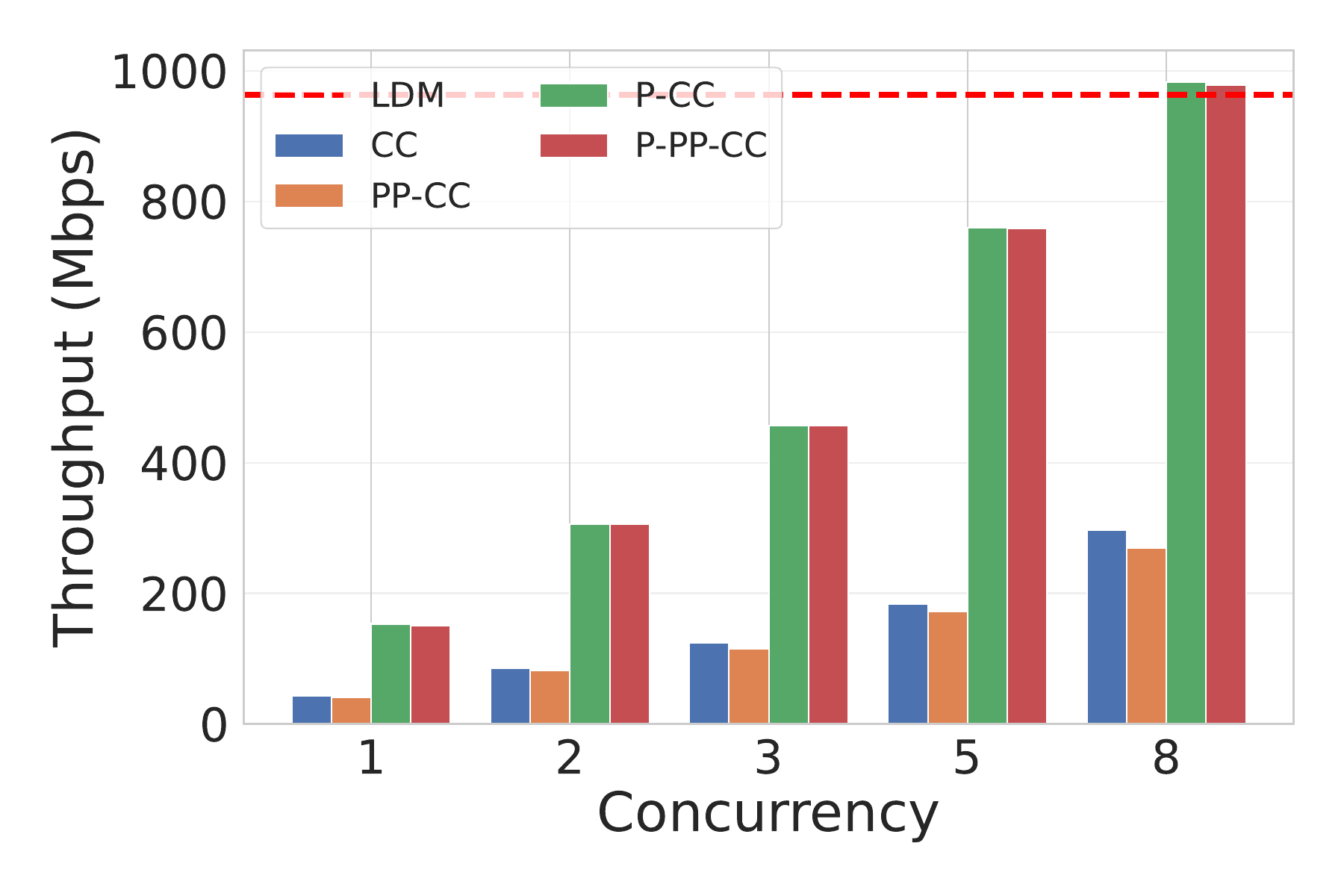}
    \\[-0.3em]
    {\small (a) 5120 files of size 1 MB}
    \vspace{2mm}
  \end{minipage}

  \begin{minipage}[b]{0.40\textwidth}
    \centering
    \includegraphics[width=\linewidth]{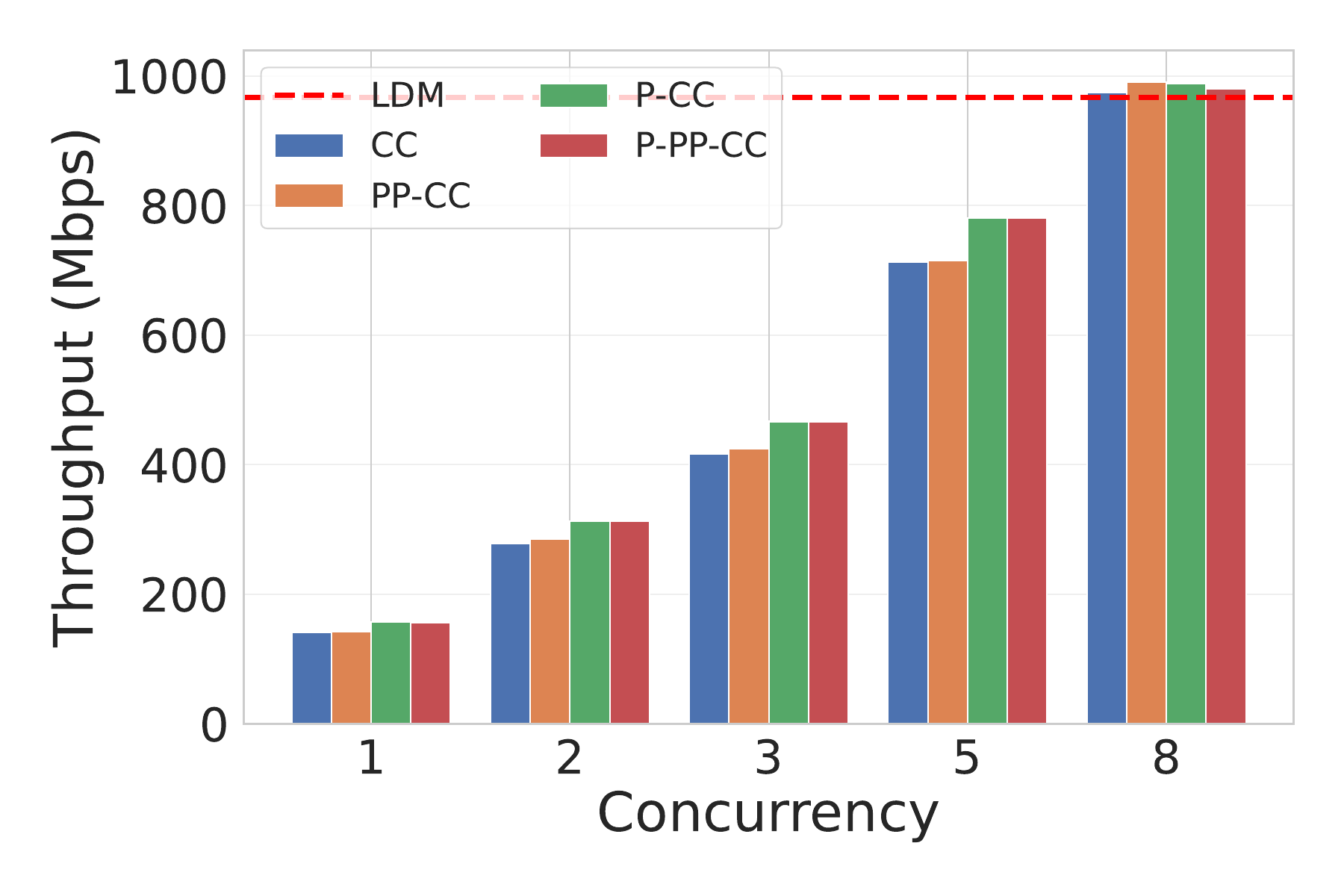}
    \\[-0.3em]
    {\small (b) 160 files of size 32 MB}
    \vspace{2mm}
  \end{minipage}

  \begin{minipage}[b]{0.40\textwidth}
    \centering
    \includegraphics[width=\linewidth]{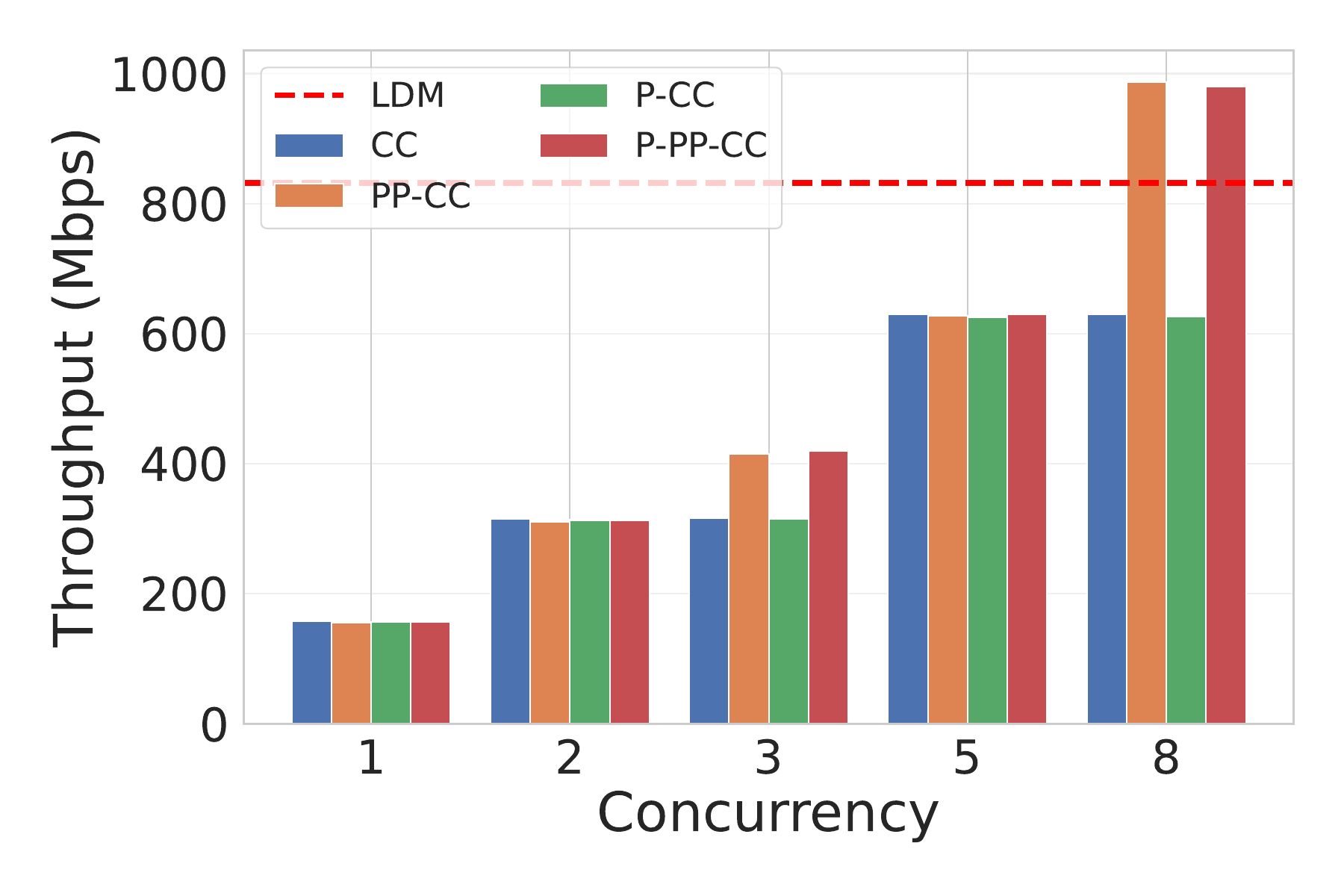}
    \\[-0.3em]
    {\small (c) 4 files of size 1 GB}
  \end{minipage}

  \caption{Ablation study on the CloudLab Clemson--Utah path for three dataset configurations. Each subplot reports throughput as a function of concurrency for different combinations of pipelining, parallelism, and concurrency.}
  \label{fig:ablation_cloudlab}
\end{figure}

\subsection{Ablation Study of Transfer Optimizations}
To understand the individual contributions of pipelining (\textit{p}), parallelism (\textit{pp}), and concurrency (\textit{cc}), we conducted an ablation study on both the Fabric (MICH--MASS) and CloudLab (Clemson--Utah) testbeds.
On the Fabric path (35 Gbps, 33 ms RTT), the bandwidth--delay product (BDP) is:
\[
35~\text{Gbps} \times 0.033~\text{s} \approx 1.155~\text{Gb} \approx 144~\text{MB}.
\]
Because each stream is manually throttled to 3 Gbps, the maximum achievable per-stream congestion window becomes:
\[
3~\text{Gbps} \times 0.033~\text{s} \approx 99~\text{Mb} \approx 12~\text{MB}.
\]

On the CloudLab path (1 Gbps, 67 ms RTT), the BDP is:
\[
1~\text{Gbps} \times 0.067~\text{s} \approx 8.4~\text{MB}.
\]
With per-stream bandwidth capped at 150 Mbps, the maximum per-stream congestion window is:
\[
150~\text{Mbps} \times 0.067~\text{s} \approx 10.05~\text{Mb} \approx 1.25~\text{MB}.
\]
We evaluated five configurations: Concurrency only (\textit{CC}), Pipelining with Concurrency (\textit{P-CC}), Parallelism with Concurrency (\textit{PP-CC}), the combined static approach (\textit{P-PP-CC}), and our fully adaptive solution (\textit{\name}), which dynamically optimizes concurrency. The red dashed line in Figures~\ref{fig:ablation_fabric} and \ref{fig:ablation_cloudlab} represents the performance of \name.

\textbf{1) Impact on 1 MB Files:}
For the dataset containing 5120 files of size 1 MB, performance is heavily constrained by control-plane latency and RTT. Since 1 MB is smaller than the maximum per-stream congestion window on both testbeds, solutions that incorporate pipelining are theoretically expected to dominate. On the Fabric testbed (Figure~\ref{fig:ablation_fabric}a), the \textit{CC} method reaches only 447 Mbps at concurrency 12, as connection setup overhead dominates the transfer time. Enabling parallelism alone (\textit{PP-CC}) performs similarly poorly ($\approx$319 Mbps).
In contrast, configurations with pipelining (\textit{P-CC} and \textit{P-PP-CC}) aggregate small files into long-lived flows, achieving roughly 9.7 Gbps, a \textbf{21x speedup} over concurrency alone. \name~correctly identifies this requirement during training and converges to an average concurrency of 8.2, achieving 8.2 Gbps. The agent considers additional threads wasteful according to its learned utility function. On Fabric, the gap between the per-stream congestion window and the file size is large, so non-pipelined methods cannot fill the pipe and their bars appear barely visible.
On the CloudLab testbed (Figure~\ref{fig:ablation_cloudlab}a), this gap is much smaller. As a result, even non-pipelined configurations show visible throughput, although pipelined variants consistently achieve the highest performance. Here as well, \name~reaches the optimal throughput without any manual tuning.

\textbf{2) Impact on 32 MB Files:}
This dataset consists of 160 files of size 32 MB, which is larger than the maximum per-stream congestion window on both testbeds. On Fabric, each file is roughly 2.7× larger than the 12 MB per-stream window; on CloudLab, each file is more than 25× larger than the 1.25 MB per-stream window. This means a single file transfer lasts long enough that slow start becomes relatively negligible for CloudLab, and pipelining is not strictly needed to hide control-plane overhead. Likewise, because the dataset contains many independent medium-sized files, parallelism is also unnecessary for both CloudLab and Fabric. For Fabric, pipelining may provide benefits because the per-file transfer time is relatively short compared to the high link capacity.
As shown in Figure~\ref{fig:ablation_fabric}b, the \textit{CC} configuration on Fabric saturates at approximately 6.4 Gbps. Parallelism alone (\textit{PP-CC}) performs slightly worse (around 5.4 Gbps) because splitting medium-sized files adds overhead without increasing effective throughput. In contrast, pipelining (\textit{P-CC}) treats the 160 files as a continuous stream, ensuring that the pipe remains full and reaching roughly 24.2 Gbps. Unlike the 1 MB case, the bars for non-pipelined variants remain visible because each 32 MB transfer is long enough to amortize some overhead.
On CloudLab (Figure~\ref{fig:ablation_cloudlab}b), all configurations perform nearly identically. Here, the per-file transfer time greatly exceeds command-channel delays, so neither pipelining nor parallelism provides measurable benefit.
\name~achieves approximately 24.7 Gbps on Fabric and 975 Mbps on CloudLab, correctly adapting to each environment.

\textbf{3) Impact on Large Files:}
The necessity of parallelism (\textit{pp}) becomes clearest when the dataset contains only a few very large files. In this experiment only, we used a chunk size of 256 MB. Thus, on Fabric (3 × 2 GB files), the system processes 3×8=24 chunks, and on CloudLab (4 × 1 GB files) it processes 4×4=16 chunks. On CloudLab, the per-stream bandwidth is capped at 150 Mbps, making it practical to fill the pipe even without pipelining. On Fabric, however, the underlying link is much faster, and slow start remains a bottleneck. This is consistent with the earlier observation that the peak throughput for 32 MB files was around 6 Gbps, whereas for large files it exceeds 10 Gbps even before parallelism is added. Therefore, Fabric benefits from both pipelining and parallelism.

As shown in Figure~\ref{fig:ablation_fabric}c, the \textit{CC} configuration peaks at 4.29 Gbps when concurrency equals the number of files (3), but throughput drops to about 3.4 Gbps at concurrency 12 because only 3 of the 12 threads have work to do; the remaining 9 sit idle and add scheduling overhead. Pipelining (\textit{P-CC}) also provides no meaningful improvement because each 2 GB file is already far larger than the per-stream congestion window. In contrast, parallelism (\textit{PP-CC}) resolves the file-count limitation by splitting each file into multiple chunks, allowing throughput to scale to roughly 13.0 Gbps. The full configuration (\textit{P-PP-CC}) achieves the highest throughput of approximately \textbf{29.0 Gbps}, showing that combining chunking with persistent connections maximizes link utilization on high-speed networks.

On CloudLab (Figure~\ref{fig:ablation_cloudlab}c), all four fixed configurations behave similarly at low concurrency (1–2). At these levels, pipelining and parallelism do not materially change throughput because each stream already transfers multiple large chunks. Differences emerge starting at concurrency 3. With parallelism enabled, each stream handles roughly five chunks, but the final remaining chunk becomes the bottleneck: once a stream finishes early, it must remain idle while one thread completes the last chunk. For the same reason, non-parallelized configurations at concurrency 2 and the parallelized configurations at concurrency 3 show nearly identical results: the tail chunk dictates completion time. This tail-effect also explains why parallelism does not outperform the others at concurrency 5; the chunk imbalance persists. At concurrency 8, however, the chunk distribution aligns more favorably, allowing parallelism-based methods to surpass all others and demonstrate their utility.

\name~matches these theoretical maxima on both testbeds. On Fabric it converges to roughly 15 threads, which is appropriate given the larger chunk pool and high link speed. On CloudLab it settles near 7 threads, but performance dips slightly due to the unavoidable imbalance caused by the final chunk. Even so, \name~still reaches a competitive throughput for this dataset, demonstrating effective adaptation to file-count limitations and chunk-level parallelism behavior.

\textbf{Summary:}
The study confirms that no single static parameter ensures performance across all scenarios. Small and medium files rely heavily on pipelining (\textit{p}) to overcome latency and BDP constraints, while large sparse datasets require parallelism (\textit{pp}) to utilize available concurrency. \name~consistently hugs the upper bound of the best-performing static configuration in every scenario, validating its hybrid adaptive design.

\begin{figure}[ht]
    \centering
    \includegraphics[width=0.45\textwidth]{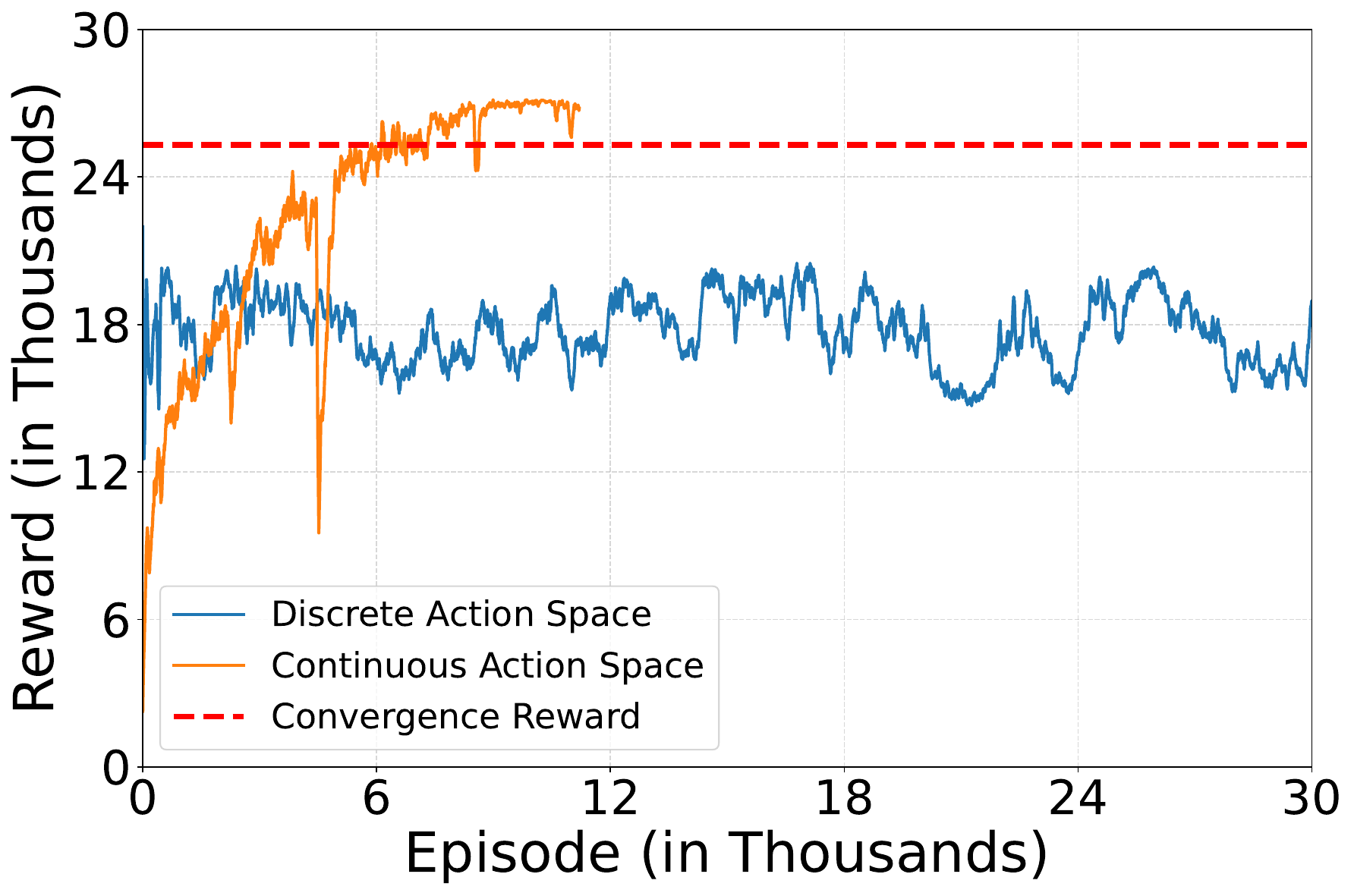}
    \caption{Training reward comparison between discrete and continuous action spaces. The continuous formulation converges faster and achieves higher stable rewards, demonstrating better suitability for high-dimensional transfer decisions.}
    \label{fig:state_space_sensitivity}
\end{figure}

\subsection{System Resource Efficiency}
To understand the computational overhead of each optimization strategy, we profiled the CPU utilization of the source node using $mpstat$ during active data transfers using Dataset A. The baseline methods, Falcon and DRL, heavily saturated the available computational resources. Both averaged approximately 71\% in system CPU usage ($\%sys$) while leaving less than 1\% of the CPU idle. This distinct lack of headroom indicates that these methods are CPU-bound, primarily due to the intense kernel-space overhead generated by repeatedly opening and closing sockets without pipelining.

In contrast, LDM demonstrated significantly higher efficiency. It maintained a system CPU usage of approximately 20\% and preserved over 56\% of the CPU capacity as idle time. This 3.5x reduction in system load confirms that the infinite pipelining strategy effectively offloads the operating system by minimizing the frequency of the TCP 3-way handshake and teardown processes. Additionally, LDM exhibited a higher soft interrupt usage ($\%soft$) of 12.2\% compared to roughly 7\% for the baselines. This metric suggests that LDM directs a greater proportion of CPU cycles toward actual packet processing and data delivery rather than administrative connection management. Finally, the offline simulator proved to be extremely lightweight, utilizing negligible system resources (0.6\%) and operating almost entirely in user space (68\%). This validates that our training environment accurately models transfer dynamics without incurring the heavy kernel costs associated with real network stacks.

\subsection{Sensitivity to State-Space Design}
We evaluate two action-space formulations for selecting concurrency. In the discrete version, the agent chooses from actions {-3, -1, 0, +1, +3}, indicating how much to increment or decrement the current concurrency. This forces the policy to move toward good values gradually, often requiring many steps to escape poor regions. In the continuous version, the agent directly predicts a real-valued concurrency level, which is then rounded to the nearest integer. This allows the policy to jump immediately to a promising concurrency level instead of relying on incremental adjustments.

Figure~\ref{fig:state_space_sensitivity} shows the training rewards for both approaches. The discrete agent learns slowly and exhibits large oscillations, while the continuous agent converges faster and reaches a higher stable reward. These results indicate that directly predicting concurrency provides a smoother optimization landscape and greater sample efficiency. For this reason,~\name adopts the continuous formulation for all experiments.

\subsection{Discussion}
Across all experiments,~\name~delivers consistent improvements over existing optimizers. The ablation study shows that pipelining is essential for small files, parallelism is required for large files, and adaptive concurrency aligns the aggregate window with the path BDP. By combining all three,~\name~automatically adapts to each dataset without manual tuning.

Generalizability results on CUBIC confirm that~\name’s gains persist even under loss-based congestion control, while the fairness study shows stable sharing in multi-tenant settings. The state-space analysis further demonstrates that continuous actions enable faster and more stable convergence. 

By addressing the limitations of static tuning and the training overhead of online learning,~\name~offers a comprehensive solution for the "Big Data" transfer challenge.

\section{Conclusion}\label{conc}
Maximizing utilization of network bandwidth in high-speed networks is a critical challenge in the era of big data. Prior work has primarily addressed this problem using a single parameter, concurrency. In our study, we demonstrate that while concurrency alone can perform adequately in certain cases, it fails to generalize across diverse datasets. To address this limitation, we introduce two additional parameters—parallelism and pipelining—that consistently outperform existing methods, achieving up to \tpimpr~Hotelsspeedup over state-of-the-art solutions. Furthermore, to reduce training overhead, we develop an offline network simulator that accelerates agent training by $2750\times$ compared to an online setting. A limitation of the current approach is that large shifts in bandwidth or RTT may require time for the agent to readapt. In such cases, retraining the simulator and policy offline is recommended. Looking ahead, our goal is to design a generalizable, plug-and-play framework that eliminates the need for retraining whenever the underlying network environment changes.

\section*{Acknowledgement}
The work in this study was supported by the NSF grant 2451376.

\bibliographystyle{plain}
\bibliography{references}

\end{document}